\documentclass[aps,prd,twocolumn,preprintnumbers,nofootinbib,superscriptaddress,amsmath]{revtex4-1}
\usepackage{graphicx}
\usepackage{amsmath}
\usepackage{amssymb}
\usepackage{amsfonts}
\usepackage{hyperref}
\usepackage{url}
\usepackage{color}
\usepackage{bm}
\usepackage{array}

\newcommand{\be}{\begin{equation}}
\newcommand{\ee}{\end{equation}}
\newcommand{\ba}{\begin{eqnarray}}
\newcommand{\ea}{\end{eqnarray}}
\newcommand{\nn}{\nonumber}

\newcommand{\pd}{\partial}

\def\lcdm{$\Lambda$CDM }

\makeatother

\begin{document}

\preprint{IFT-UAM/CSIC-20-80}

\title{Cosmological constraints on nonadiabatic dark energy perturbations}

\author{Rub\'{e}n Arjona}
\email{ruben.arjona@uam.es}
\author{Juan Garc\'{i}a-Bellido}
\email{juan.garciabellido@uam.es}
\author{Savvas Nesseris}
\email{savvas.nesseris@csic.es}

\affiliation{Instituto de F\'isica Te\'orica UAM-CSIC, Universidad Auton\'oma de Madrid,
Cantoblanco, 28049 Madrid, Spain}


\date{\today}

\begin{abstract}
The exact nature of dark energy is currently unknown and its cosmological perturbations, when dark energy is assumed not to be the cosmological constant, are usually modeled as adiabatic. Here we explore the possibility that dark energy might have a nonadiabatic component and we examine how it would affect several key cosmological observables. We present analytical solutions for the growth rate and growth index of matter density perturbations and compare them to both numerical solutions of the fluid equations and an implementation in the Boltzmann code CLASS, finding that they all agree to well below one percent. We also perform a Monte Carlo analysis to derive constraints on the parameters of the  nonadiabatic component using the latest cosmological data, including the temperature and polarization spectra of the Cosmic Microwave Background as observed by Planck, the Baryon Acoustic Oscillations, the Pantheon type Ia supernovae compilation and lastly, measurements of Redshift Space Distortions (RSD) of the growth rate of matter perturbations. We find that the amplitude of the nonadiabatic pressure perturbation is consistent with zero within $1\sigma$. Finally, we also present a new, publicly available, RSD likelihood for MontePython based on the ``Gold 2018'' growth rate data compilation.
\end{abstract}
\maketitle

\section{Introduction \label{sec:intro}}
Recent observations of type Ia supernovae (SnIa) at the end of the previous century have indicated that on cosmological scales the Universe is undergoing a phase of accelerated expansion, usually attributed to the cosmological constant $\Lambda$ \cite{Riess:1998cb,Perlmutter:1998np}. Since then, this finding has been confirmed via a plethora of different observations something which, in conjunction with theoretical developments, has led to the creation of a robust description of the evolution of the Universe on cosmological scales within the framework of general relativity (GR). This paradigm is known as the standard $\Lambda$ cold dark matter model ($\Lambda$CDM) and it contains just six free parameters, which describe the dark energy (DE) and matter  contents of the cosmos. Currently, the \lcdm model is our best phenomenological description of the data \cite{Aghanim:2018eyx}.

However, since the first detection of DE, several alternatives to the \lcdm model have also been developed, which roughly fall under the umbrella of two main categories. First, there are the so-called Modified Gravity (MG) models \cite{Clifton:2011jh}, which assume that GR is  modified on large scales, the so-called Infrared (IR) modifications, in order to accommodate current observations \cite{Bertschinger:2011kk}. However, certain modifications of GR are fraught with difficulties, such as the Ostrogradsky instability, that arises when a nondegenerate Lagrangian with time derivatives higher than second order, leads to an unstable Hamiltonian \cite{Woodard:2006nt,Motohashi:2014opa}. Furthermore, several tests with cosmological data seem to be in very good agreement with GR \cite{Bertotti:2003rm,Reyes:2010tr,PhysRevLett.116.221101,PhysRevLett.121.231101,PhysRevLett.121.231102,Ishak:2018his,Luna:2018tot,Basilakos:2018arq,Perez-Romero:2017njc, Basilakos:2017rgc,Nesseris:2017vor,Basilakos:2016nyg}. 

The second category of theories that are serious contenders to \lcdm are DE models \cite{Copeland:2006wr} with as yet unobserved scalar fields that dominate over the other matter species at late times, while at the same time, avoiding fine-tuning  \cite{Ratra:1987rm,ArmendarizPicon:2000dh}. Most of these DE models also exhibit perturbations, which will affect the large scale structure (LSS) of the Universe; however they tend to be subdominant at late times and on scales of interest. As a result, in order to constrain the cosmological parameters to a percent level and discriminate between the various theories, DE perturbations should be well understood as they are expected to play an important role in the near future \cite{DeDeo:2003te,Bean:2003fb,Kunz:2006wc,Kunz:2006ca, Sapone:2009mb}.

These two categories seem at a first glance quite dissimilar, however it is possible to unify them within the same framework. One way to do this is to map the MG models, to linear order, to some DE fluid via the effective fluid approach. Then, MG models can be interpreted as DE fluids described by an equation of state $w(a)$, a pressure perturbation $\delta P(k,a)$, and an anisotropic stress $\sigma(k,a)$ \cite{Kunz:2006ca,Pogosian:2010tj,Arjona:2018jhh,Arjona:2019rfn,Capozziello:2005mj,Capozziello:2006dj,Capozziello:2018ddp}. Hence, the evolution of the background is determined by $w(a)$, while the evolution of the perturbations is governed by $\delta P(k,a)$ and $\sigma(k,a)$, both of which are time and scale-dependent. In this case however, the effective fluid DE pressure perturbation $\delta P(k,a)$ could also be interpreted as containing both an adiabatic and a nonadiabatic contribution, as we will see later on in Sec.~\ref{sec:theory}. 

On the other hand, the presence of DE anisotropic stress has the interesting side-effect that the DE sound speed $c_{\textrm{s,DE}}^2$ can in general be negative, without sacrificing the overall stability of the perturbations. This is true as long as the effective sound speed, which is the sum of the DE sound speed and the anisotropic stress, is always positive \cite{Cardona:2014iba}. Moreover, it can be shown that a varying adiabatic sound speed of DE perturbations can mimic anisotropic stresses \cite{Koivisto:2005mm,Mota:2007sz}. 

In this paper we will consider a holistic approach and also consider nonadiabatic DE perturbations, motivated by the following reasons. First, in Ref.~\cite{Arjona:2020kco} it was shown with a machine learning approach, based on the Genetic Algorithms, that current data seem to give hints for the existence of DE anisotropic stress, thus going beyond simple DE models within GR. This could also leave open the possibility for a nonadiabatic DE component, as then the DE component could originate from a higher energy model, usually of the MG type. Second, the previous observation is crucial since, as mentioned earlier and will be seen in detail in the following sections, when MG models are described by the effective fluid approach, equivalently they can also be modeled as a DE fluid with a nonadiabatic component. Hence, we conclude that a nonadiabatic DE component could arise naturally in a wide class of models.

Finally, Primordial Black Holes (PBH) can be a significant component of Dark Matter~\cite{Garcia-Bellido:2017fdg} and give rise to entropy perturbations at early times on very small scales. They grow like isocurvature energy density perturbations and may eventually generate a significant component on large scales~\cite{Tada:2015noa}. Note that PBH as dark matter behaves as an adiabatic component on very large scales, since it follows the large scale curvature perturbations just like baryons and photons. It is only on small scales that it has an isocurvature component, which is also highly non-Gaussian and can grow to become relevant at late times, around vacuum energy domination. While the PBH entropy perturbations happen on very different scales from those of DE, this clearly provides another mechanism for giving rise to a nonadiabatic component in the dark sector.

Here we consider the effects of the nonadiabatic DE perturbations on the LSS of the Universe, as the latter is directly affected by the underlying gravitational theory, something which allows us to easily search for deviations from GR. A main probe of LSS is the matter density perturbations, which in linear theory can be parameterized through the growth parameter $\delta_m=\frac{\delta \rho_m}{\bar{\rho}_m}$ and the growth rate $f\equiv \frac{d \ln \delta_m}{d\ln a}$, which is the former's logarithmic derivative while $\bar{\rho}_m$ is the background matter density and $\delta \rho_m$ its perturbation to linear order. The growth rate can also be parameterized via the growth index $\gamma$ parameter \cite{Wang:1998gt}, which in the \lcdm model is equal to $\gamma\simeq 6/11$, hence making it easier to look for deviations from GR. The growth index is defined as the exponent of the growth rate $f(z)=\Omega^{\gamma}_m(z)$ and, as in the \lcdm model the growth rate is scale-invariant on large scales, this makes $\gamma$ a useful discriminator of DE models \cite{Franco:2019wbj}.

One of the main advantages of the growth rate is that it encodes information about how gravity affects the LSS, as the latter requires only linear physics, which is well understood. This means the growth can be a particularly useful probe \cite{Akrami:2018vks}. Similarly, the growth index can help discriminate models both between DE and MG (see Ref.~\cite{Basilakos:2017rgc,Perez-Romero:2017njc}) and between \lcdm \cite{Luna:2018tot} and MG models that are fully degenerate at the background level \cite{Nesseris:2013fca,Arjona:2018jhh,delaCruzDombriz:2006fj,Multamaki:2005zs,Pogosian:2007sw}.

Some of the first constraints on the sound speed of DE were reported in Ref.~\cite{Hannestad:2005ak} by using WMAP data. However, given the data at the time, no significant sensitivity on the adiabatic sound-speed was reported. On  the other hand, nonadiabatic perturbations were studied within the context of a decaying vacuum cosmology in  Ref.~\cite{Zimdahl:2011ae}, where they were found to only have an effect on larger scales. Constraints on nonadiabatic DE models using only growth RSD data were reported in Ref.~\cite{Velten:2017mtr} which used a particular parameterization for the nonadiabatic DE perturbations based on a linear combination of the intrinsic and entropy perturbations $\Gamma(a)$ and $S(a)$ \cite{Velten:2017mtr}. Using a conjoined analysis of the $f\sigma_8$ and $H(z)$ data no deviations from \lcdm were found. Another similar analysis with only growth RSD data was done in Ref.~\cite{Zimdahl:2019pqg}, which did not find any deviations from the standard cosmological model. Finally, a related approach in the search of primordial entropy perturbations was presented in Ref.~\cite{Romano:2018frb} and was constrained by the Cosmic Microwave Background (CMB) data  in Ref.~\cite{Rodrguez:2020hot}.

In the next sections we will present a broader approach by considering a general ansatz for the nonadiabatic DE perturbations and we will use the latest cosmological data, including Planck 18, BAO and RSD measurements to constrain its model parameters. The structure of our paper is as follows. In Sec.~\ref{sec:theory} we present the theoretical background of our analysis and a realistic parameterization for the nonadiabatic DE pressure perturbations, along with analytic solutions for the growth of matter density perturbations and the growth index $\gamma$, while in Sec.~\ref{sec:class} we compare our numerical and analytical solutions against an implementation of the nonadiabatic perturbations in the Boltzmann code CLASS. In Sec.~\ref{sec:mcmc} we present our results from a Monte Carlo Markov Chain (MCMC) analysis using the latest cosmological data, while in Sec.~\ref{sec:conclusions} we discuss our conclusions. Finally, in Appendix~\ref{sec:appendix1} we present an implementation of the redshift space distortions (RSDs) likelihood for MontePython.

\section{Theory \label{sec:theory}}
We will consider a spatially flat universe and assume that the scalar perturbations of the metric can be described by the perturbed Friedmann-Robertson-Walker metric in the conformal Newtonian gauge
\be
ds^2=a^2\left[-(1+2\psi)d\tau^2+(1-2\phi)dx_idx^i\right],
\ee
where $a=a(\tau)=\frac{1}{1+z}$ is the scale factor, $z$ is the redshift and $d\tau=dt/a$ is the conformal time in terms of the cosmic time $t$.

We assume that a DE fluid is responsible for the accelerated expansion of the universe and that its background evolution can be described by an equation of state $w=\bar{P}/\bar{\rho}$, while its fluctuations can be described by a pressure perturbation $\delta P$ and anisotropic stress $\sigma$. The energy momentum tensor of the fluid can be written as
\be
T^\mu_\nu = P \delta^\mu_\nu+(\rho+P)U^\mu U_\nu,
\ee
where the overhead bar $\bar{\rho}$ denotes a background quantity, $U^\mu\equiv dx^\mu/\sqrt{-ds^2}$ is the four velocity, given to linear order by $U^\mu\simeq \frac1{a}\left(1-\psi,u^i\right)$ for $u^i=dx^i/d\tau$ and the density and pressure include both background and perturbations, i.e. $\rho=\bar{\rho}+\delta \rho$ and $P=\bar{P}+\delta P$. The components of the energy momentum tensor are then given by
\ba
T^0_0&=&-(\bar{\rho}+\delta \rho),\\
T^0_i&=& (\bar{\rho}+\bar{P})u_i,\\
T^i_j&=& (\bar{P}+\delta P)\delta^i_j+\Sigma^i_j,
\ea
where $\Sigma^i_j$ is the anisotropic stress tensor, which is traceless $\Sigma^i_i=0$ and can also be written via the $\sigma$ parameter as   $(\bar{\rho}+\bar{P})\sigma\equiv-(\hat{k}^i\hat{k}_j-\frac13 \delta^i_j)\Sigma^j_i$.

The evolution equations of the fluid variables $\delta=\frac{\delta \rho}{\bar{\rho}}$ and velocity of the DE fluid $\theta=ik^j u_j$ can be found by the conservation of the energy momentum tensor $T^{\mu\nu}{}_{;\nu}=0$ and are given by  \cite{Ma:1995ey,Sapone:2009mb}:
\ba
\dot{\delta}&=&-(1+w)\left(\theta-3\dot{\phi}\right)-3\mathcal{H}\left(\frac{\delta P}{\bar{\rho}}-w\delta\right), \label{eq:eqd}\\
\dot{\theta}&=&-\mathcal{H} (1-3w)\theta-\frac{\dot{w}}{1+w}\theta+\frac{\delta P/\bar{\rho}}{1+w}k^2-k^2\sigma+k^2\psi,~~~\label{eq:eqth}
\ea
where $\mathcal{H}\equiv\frac{\dot{a}}{a}$ is the conformal Hubble parameter and $k$ is the wavenumber of the Fourier mode of the perturbations, which in GR are decoupled.

In general, is is most convenient to describe the DE pressure perturbation in the rest frame $\hat{\delta P}$, denoted here by a hat, which is defined as the frame where the fluid is at rest, i.e. $\hat{\theta}=0$. Then, the pressure perturbation in the rest frame can be expressed in terms of the energy density $\rho$ and entropy $S$ as $\hat{P}=\hat{P}(\rho,S)$ as \cite{Christopherson:2008ry}
\be
\hat{\delta P}=\left.\frac{\hat{\pd P}}{\pd \rho}\right|_S\hat{\delta \rho}+\left.\frac{\hat{\pd P}}{\pd S}\right|_\rho \hat{\delta S},
\ee
where the DE density and entropy perturbations at the rest frame are given by $\hat{\delta \rho}$ and $\hat{\delta S}$ respectively. In principle, the nonadiabatic contribution may come from some internal degrees of freedom, as for example happens in the quintom model \cite{Kunz:2006wc}. We can straight-forwardly identify the DE rest frame sound speed as
\be
\hat{c}_s^2\equiv \left.\frac{\hat{\pd P}}{\pd \rho}\right|_S,
\ee
which is equal to one for quintessence, but is in the range $\hat{c}_{s}^{2}\in[0,1]$ for k-essence or other models \cite{Amendola:2015ksp}. For modified gravity models it can even be negative, in which case one would presume that a negative value would cause instabilities in the perturbations, unless there is anisotropic stress to stabilize them \cite{Cardona:2014iba}.

We can now decompose the pressure perturbation in terms of the sound speed $\hat{c}_s^2$ and a nonadiabatic part $\hat{\delta P}_{\textrm{nad}}$ as
\be
\hat{\delta P}=\hat{c}_s^2 \bar{\rho} \hat{\delta}+\hat{\delta P}_{\textrm{nad}},\label{eq:dPhat1}
\ee
where both quantities are defined in the DE rest frame and the nonadiabatic contribution at the rest frame can be identified as
\be
\hat{\delta P}_{\textrm{nad}}=\left.\frac{\hat{\pd P}}{\pd S}\right|_\rho \hat{\delta S}.
\ee

In order to use the aforementioned expressions for the pressure perturbation in any other frame besides the DE rest frame, we have to change gauge by considering a general coordinate transformation between the hatted (DE rest frame) and the unhatted (general) frame \cite{Ma:1995ey,Kunz:2006wc}:
\be
x^\mu =\hat{x}^\mu +d^\mu,
\ee
where $d^\mu=(\alpha(\vec{x},\tau),\vec{\nabla} \beta(\vec{x},\tau)+\vec{\epsilon}(\vec{x},\tau))$, for some functions $\alpha$, $\beta$ and $\epsilon$. Then, the perturbation variables transform as \cite{Ma:1995ey}
\ba
\delta &=& \hat{\delta}-\alpha \frac{\dot{\bar{\rho}}}{\bar{\rho}}, \label{eq:trans1}\\
\theta &=& \hat{\theta}-\alpha k^2, \label{eq:trans2}\\
\delta P &=& \hat{\delta P}-\alpha \dot{\bar{P}},\label{eq:trans3}
\ea
where in the rest frame we have that $\hat{\theta}=0$. We can use Eq.~\eqref{eq:trans2} to eliminate $\alpha$, as $\hat{\theta}=0$, thus finding from Eq.~\eqref{eq:trans3}
\be
\delta P= \hat{\delta P}-3\mathcal{H}c_a^2 \bar{\rho}\frac{(1+w)\theta}{k^2},\label{eq:gaugeco}
\ee
where $c_a^2=\frac{\dot{\bar{P}}}{\dot{\bar{\rho}}}=w-\frac{\dot{w}}{3\mathcal{H} (1+w)}$ is the so-called adiabatic sound speed and we have used the background conservation equation
\be
\dot{\bar{\rho}}+3\mathcal{H}(1+w)\bar{\rho}=0.\label{eq:backrho}
\ee
Using Eqs.~\eqref{eq:dPhat1}, \eqref{eq:trans1} and \eqref{eq:backrho} in Eq.~\eqref{eq:gaugeco} we can write the pressure perturbation in any gauge as 
\be
\delta P= \hat{c}_s^2 \bar{\rho} \delta+\hat{\delta P}_{\textrm{nad}}+3\mathcal{H}\left(\hat{c}_s^2-c_a^2\right) \bar{\rho}\frac{(1+w)\theta}{k^2},\label{eq:gaugeco1}
\ee
which is in agreement with Ref.~\cite{Kunz:2006wc}. Thus, our final expressions for the evolution equations for the DE perturbations in the conformal Newtonian gauge are given by
\ba
\dot{\delta}_\textrm{DE}&=&-(1+w)\left(\theta_\textrm{DE}-3\dot{\phi}\right)-3\mathcal{H}\left(\hat{c}_s^2-w\right)\delta_\textrm{DE}
\nn\\
&-&9\mathcal{H}^2\left(\hat{c}_s^2-c_a^2\right)\frac{(1+w)\theta_\textrm{DE}}{k^2}-3\mathcal{H}\frac{\hat{\delta P}_{\textrm{nad}}}{\bar{\rho}}, \label{eq:eqd2}\\
\dot{\theta}_\textrm{DE}&=&-\mathcal{H} (1-3\hat{c}_s^2)\theta_\textrm{DE}+\frac{k^2\hat{c}_s^2}{1+w}\delta_\textrm{DE}-k^2\sigma+k^2\psi\nn \\
&+&\frac{\hat{\delta P}_{\textrm{nad}}/\bar{\rho}}{1+w}k^2.\label{eq:eqth2}
\ea
Compared to Refs.~\cite{Ma:1995ey} and \cite{Sapone:2009mb}, the last terms in Eqs.~\eqref{eq:eqd2} and \eqref{eq:eqth2} are new. The latter, ignoring any nonadiabatic contributions, are commonly used in the Boltzmann codes to model the behavior of the DE perturbations. In order to include them in the aforementioned codes, we will henceforth assume that the DE fluid in the rest frame also has a nonadiabatic component $\hat{\delta P}_{\textrm{nad}}$.

This extra component however, can in principle destabilize the perturbations. To demonstrate this, we follow  Ref.~\cite{Cardona:2014iba} and we eliminate $\theta$ from Eqs.~\eqref{eq:eqd2}-\eqref{eq:eqth2}, resulting in a second order equation for the growth of DE perturbations $\delta_\textrm{DE}$:
\ba
\ddot{\delta}_\textrm{DE}&+&(\cdots)\dot{\delta}_\textrm{DE}+(\cdots)\delta_\textrm{DE} = \nn \\
&=& -k^2\left((1+w)\psi+\hat{c}_s^2\delta_\textrm{DE}+\hat{\delta P}_{\textrm{nad}}/\bar{\rho} -\frac23\pi\right)+\cdots,~~~ \nn\\
\ea
where the dots $(\cdots)$ indicate the presence of complicated expressions and we have redefined the anisotropic stress parameter of the DE fluid as $\pi\equiv \frac32 (1+w)\sigma$. Here we focus solely on the last $k^2$ term, which as discussed
in Ref.~\cite{Kunz:2006wc}, it will act as a source driving the perturbations. However, since the potential scales as $\psi\sim1/k^2$ in matter domination, the only terms that matter are the sound speed, the nonadiabatic perturbation and the anisotropic stress. Therefore, we can define an effective sound speed as
\be
c_{s,\textrm{eff}}^2=\hat{c}_s^2+\frac{\hat{\delta P}_{\textrm{nad}}}{\bar{\rho}\;\delta_\textrm{DE}} -\frac23\pi/\delta_\textrm{DE},
\ee
which has to be positive for the perturbations to be stable at all scales.

In order to solve Eqs.~\eqref{eq:eqd2} and \eqref{eq:eqth2}, we need to choose a parameterization for the DE nonadiabatic pressure perturbations, something which is non-trivial in general without using an underlying model. Hence, in order to keep our results general enough, in what follows we will attempt to motivate an ansatz for the evolution of the pressure perturbations, by using a realistic $f(R)$ to determine the behavior of $\delta P$ at early and late times. As dark energy is only expected to dominate at late times, the initial conditions will not affect our results, but we will still discuss them for completeness. In particular, here we will consider a case which is motivated by the effective fluid approach of Refs.~\cite{Arjona:2018jhh,Arjona:2019rfn} and as an example we will consider the designer $f(R)$ model, see Ref.~\cite{Arjona:2018jhh}, which is constructed so that the background expansion corresponds exactly to \lcdm but to linear order, it can have perturbations \cite{Nesseris:2013fca}. This implies that $w=-1$ and from Eq.~\eqref{eq:gaugeco1} we have that for the designer $f(R)$ model
\be
\frac{\hat{\delta P}_{\textrm{nad,des}}}{\bar{\rho}}=\frac{\delta P}{\bar{\rho}}-\hat{c}_s^2 \delta_\textrm{DE}-3\mathcal{H}\left(\hat{c}_s^2-c_a^2\right) \frac{V_\textrm{DE}}{k^2},
\ee
where $\frac{\delta P}{\bar{\rho}}$ and $\delta_\textrm{DE}$ are given by Eqs.~(42) and (43) of Ref.~\cite{Arjona:2018jhh}, $V_\textrm{DE}=(1+w)\theta_\textrm{DE}$, while $\hat{c}_s^2=1$ for $f(R)$. Note that for this model, in general we have $V_\textrm{DE}\neq 0$ even if $w=-1$ \cite{Arjona:2018jhh}.

\begin{figure}[!t]
\centering
\includegraphics[width=0.49\textwidth]{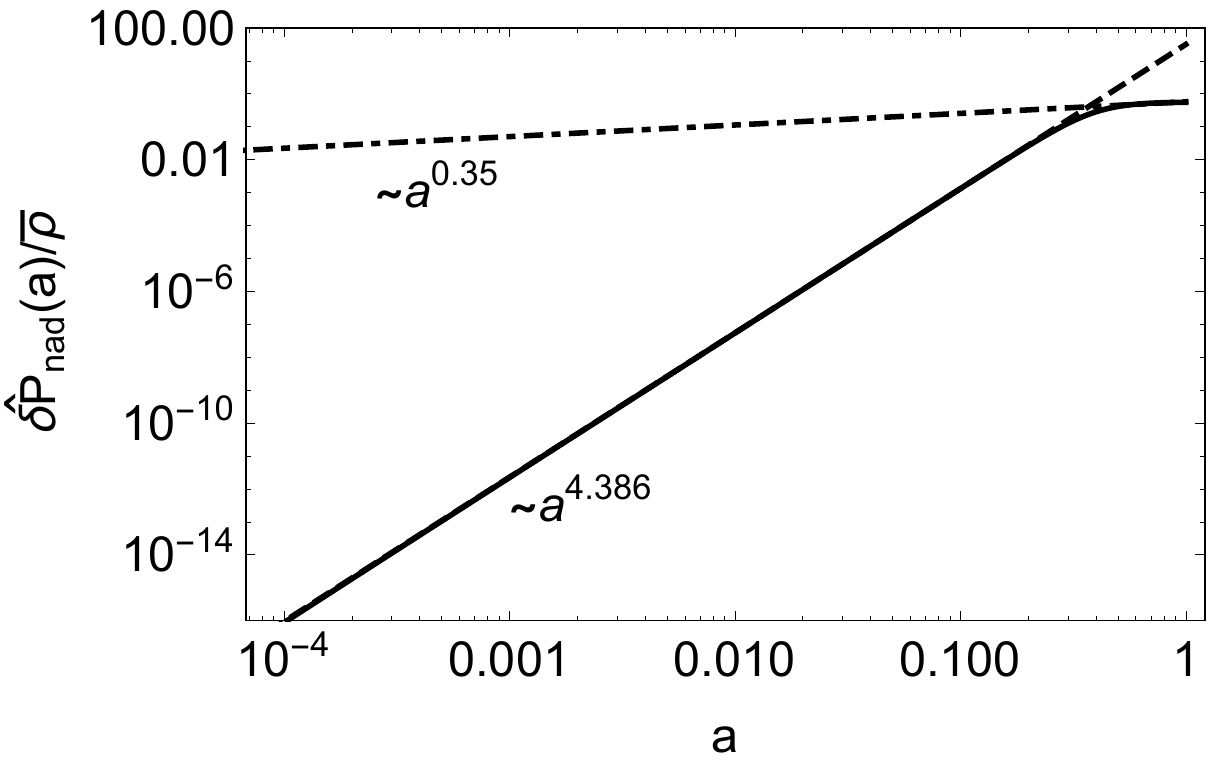}
\caption{The evolution of $\frac{\hat{\delta P}_{\textrm{nad}}}{\bar{\rho}}$ for the  designer $f(R)$ model obtained using the effective fluid approach of Ref.~\cite{Arjona:2018jhh}, for $\Omega_{m0}=0.3$, $f_{R0}=-10^{-4}$, $k=300\,H_0$ and $w=-1$. As seen, at different times the nonadiabatic component evolves as a power law of the form $\frac{\hat{\delta P}_{\textrm{nad}}}{\bar{\rho}} \sim c_0 a^n k^2$. The solid black line corresponds to the prediction from the designer $f(R)$ model, while the dashed and dot-dashed lines correspond to the asymptotic limits at early and late times respectively.
\label{fig:fRdesigner}}
\end{figure}

We plot this function for the designer $f(R)$ model for $\Omega_{m0}=0.3$, $f_{R0}=-10^{-4}$  and $w=-1$ in Fig.~\ref{fig:fRdesigner}, where we see that at both early and late times, the nonadiabatic component evolves as a power law of the form $\hat{\delta P}_{\textrm{nad}}/\bar{\rho} \simeq c_0 a^n k^2/H_0^2$. Specifically, we find that \cite{Arjona:2018jhh}
\ba
n&=&\frac94+\frac{\sqrt{73}}{4}\nn\\
&\simeq&4.386, \\
c_0&=& -\frac{5+\sqrt{73}}{36}g(\Omega_{m0})f_{R0},
\ea
where 
\be
g(\Omega_{m0})\simeq\frac{\Omega_{m0}^{-\frac{17}{12}-\frac{\sqrt{73}}{12}}}{\, _2F_1\left[\frac{\sqrt{73}+5}{12}, \frac{\sqrt{73}+11}{12};\frac{\sqrt{73}+6}{6};1-\Omega_{m0}\right]}.
\ee
Inspired by this functional form, in what follows we will assume the rather general ansatz
\ba
\frac{\hat{\delta P}_{\textrm{nad}}}{\bar{\rho}}&=&c_0\;a^n k^2/H_0^2,\label{eq:Pansatz}
\ea
where $(c_0,n)$ are parameters to be determined; however the exponent $n$ has to be positive so as to ensure the nonadiabatic DE perturbation vanishes at early times, and thus we will assume the prior $n\in(0,\infty)$. In the next sections we will present constraints on the parameters $(c_0,n)$ in the case of $w=$const and of no DE anisotropic stress ($\sigma=0$).

\subsection{The initial conditions\label{sec:ICsnonad}}

Here will now discuss the initial conditions for the DE perturbations in both gauges and in two different regimes, in matter and radiation domination. First, we consider the initial conditions in the conformal Newtonian gauge in matter domination, for which we follow Ref.~\cite{Sapone:2009mb}. In a similar vein we consider two regimes: 1) the DE perturbations are larger than the sound horizon, $k\ll aH/\hat{c}_s$ or equivalently that $\hat{c}_s^2=0$; 2) the small scales solutions $k\gg aH/\hat{c}_s$, which implies that the terms scaling as $k^2$ dominate of over the rest.

In any case, the initial conditions for matter and the potential (assuming no anisotropic stress) in matter domination are unchanged and given by \cite{Sapone:2009mb}
\ba
\delta_m(a)&=& \delta_0 \left(a+\frac{3H_0^2\Omega_{m0}}{k^2}\right) ,\\
V_m(a)&=&-\delta_0 H_0 \sqrt{\Omega_{m0}}\; a^{1/2},\\
\phi &=& -\frac32\delta_0\frac{H_0^2\Omega_{m0}}{k^2},
\ea
where $\delta_0$ is a normalization set at early times from inflation, while $V_i\equiv (1+w_i)\theta_i$.

In the first case $(k\ll aH/\hat{c}_s)$ we find that the initial conditions for the DE density and velocity perturbations are given by
\ba
\delta_{\textrm{DE}}(a)&=& \delta_0 (1+w) \left(\frac{a}{1-3w}+\frac{3H_0^2\Omega_{m0}}{k^2}\right)\nn \\
&-&\frac{c_0 k^2 a^n \left(6+\frac{9}{n+3 w}-\frac{2 a k^2/H_0^2}{n \Omega_{m0}-3 \Omega_{m0} w+\Omega_{m0}}\right)}{H_0^2 (2 n+3)},~~~\\
V_{\textrm{DE}}(a)&=&-\delta_0(1+w) H_0\sqrt{\Omega_{m0}}\;a^{1/2}\nn \\
&+&\frac{c_0 k^4 a^{n+\frac{1}{2}}}{H_0^3 \left(n+\frac{3}{2}\right) \sqrt{\Omega_{m0}}} .
\ea

In the second case $(k\gg aH/\hat{c}_s)$ we find that the initial conditions for the DE density and velocity perturbations are given by
\ba
\delta_{\textrm{DE}}(a)&=& \frac32 (1+w) \delta_0\frac{H_0^2 \Omega_{m0}}{\hat{c}_s^2 k^2}-\frac{c_0 k^2 a^n}{\hat{c}_s^2 H_0^2},\label{eq:dDEICMD}\\
V_{\textrm{DE}}(a)&=&-\frac92(1+w) (\hat{c}_s^2-w)\frac{H_0^3\Omega_{m0}^{3/2}}{\hat{c}_s^2\;k^2}a^{-1/2}\nn \\
&+& \frac{c_0 k^2 \sqrt{\Omega_{m0}} a^{n-\frac{1}{2}} (n-3 w)}{\hat{c}_s^2 H_0}\bigg[1-\frac{9 H_0^2 \Omega_{m0} (\hat{c}_s^2-w)}{a k^2}\nn \\
&+& \frac{81 H_0^4 \Omega_{m0}^2 (\hat{c}_s^2-w)^2}{a^2 k^4}\bigg].
\ea
Note that in the previous sets of equations, the dark energy perturbations have non-vanishing values, even for $w=-1$. This is clearly a smoking gun signal for modified gravity, as the usual dark energy perturbations within GR exactly vanish for the cosmological constant $(w=-1)$. Finally, we find that in both cases the last terms containing $c_0$, are new compared to Ref.~\cite{Sapone:2009mb} and correspond to the contribution of the nonadiabatic term.

For the simpler case of a constant adiabatic DE sound-speed $\hat{c}_s^2$, the initial conditions in the synchronous gauge in radiation domination where first derived in Ref.~\cite{Ballesteros:2010ks} as a series expansion in terms of $k\tau$. Here we generalize this approach by also considering the nonadiabatic pressure perturbation and we follow Refs.~\cite{Ma:1995ey,Ballesteros:2010ks}. Since we have to expand in terms of $k\tau$ we find that in this case it is more convenient to consider the different regimes for the index $n$ of the power law of our ansatz given by Eq.~\eqref{eq:Pansatz}. Specifically, as we have already mentioned, $n$ has to be positive in order for the nonadiabatic pressure perturbation to vanish at early times, so we will consider the regimes $n\in(0,1)$, $n\in[1,2)$,  $n\in[1,2)$ and $n\geq3$, since then the scalar factor dominates differently at early times. 

Then, by expanding the Einstein and fluid equations in terms of $k\tau$, following Refs.~\cite{Ma:1995ey,Ballesteros:2010ks}, we find the initial conditions for the DE density $\delta_{\textrm{DE}}$ and velocity $\theta_{\textrm{DE}}$ perturbations for $n\in(0,1)$
\ba
\delta_{\textrm{DE}}(a)&=& \frac{\delta_0 (3 \hat{c}_s^2-4) (k\tau)^2 (w+1)}{6\hat{c}_s^2-12 w+8}\nn \\
&+&\frac{c_0 k^2 }{4 H_0^2 (3 \hat{c}_s^2-6 w+4) (\hat{c}_s^2-w)} \nn \\
&\cdot& \bigg[4 \left(w \left((k \tau)^2-9 w+12\right)-4\right)\nn \\
&-&3 \hat{c}_s^2 \left(\left((k \tau)^2-6\right) w+4\right)\bigg],\label{eq:ICs1}\\
\theta_{\textrm{DE}}(a)&=& -\frac{\delta_0 \hat{c}_s^2 k (k\tau)^3}{6 \hat{c}_s^2-12 w+8}\nn \\
&+&\frac{c_0 k^3 (k \tau) w \left(\hat{c}_s^2 \left((k \tau)^2-6\right)+12 w-8\right)}{4 H_0^2 (w+1) (3 \hat{c}_s^2-6 w+4) (\hat{c}_s^2-w)}.
\ea
For  $n\in[1,2)$ we have that 
\ba
\delta_{\textrm{DE}}(a)&=& \frac{\delta_0 (3 \hat{c}_s^2-4) (k\tau)^2 (w+1)}{6\hat{c}_s^2-12 w+8}\nn \\
&+&\frac{3 a c_0 k^2 (w-1)}{H_0^2 (2 \hat{c}_s^2-3 w+1)},\\
\theta_{\textrm{DE}}(a)&=& -\frac{\delta_0 \hat{c}_s^2 k (k\tau)^3}{6 \hat{c}_s^2-12 w+8}\nn \\
&+&\frac{a c_0 k^3 (k \tau) (3 w-1)}{3 H_0^2 (w+1) (-2 \hat{c}_s^2+3 w-1)}.
\ea
For  $n\in[1,2)$ we have that 
\ba
\delta_{\textrm{DE}}(a)&=& \frac{\delta_0 (3 \hat{c}_s^2-4) (k\tau)^2 (w+1)}{6\hat{c}_s^2-12 w+8}\nn \\
&+&\frac{3 a^2 c_0 k^2 (3 w-4)}{H_0^2 (6 \hat{c}_s^2-12 w+8)},\\
\theta_{\textrm{DE}}(a)&=& -\frac{\delta_0 \hat{c}_s^2 k (k\tau)^3}{6 \hat{c}_s^2-12 w+8}\nn \\
&+&\frac{a^2 c_0 k^3 (k \tau) (3 w-2)}{2 H_0^2 (w+1) (-3 \hat{c}_s^2+6 w-4)},\label{eq:ICs2}
\ea
while for $n\geq3$ the contribution from the nonadiabatic pressure perturbation of Eq.~\eqref{eq:Pansatz} is subdominant and we recover the results of Ref.~\cite{Ballesteros:2010ks}.

\subsection{Approximate solutions and the growth index \label{sec:analytic}}
Here we present analytic solutions to the evolution equations \eqref{eq:eqd2} and \eqref{eq:eqth2}, as well as analytic expressions for the growth index $\gamma$ at late times. We note that the forthcoming approximations are only used to gain insight and intuition on the effects of the nonadiabatic term on the growth and the LSS and are not used in CLASS or the MCMC analysis later on, for which we solve the corresponding equations numerically.

One way to determine how the nonadiabatic DE pressure perturbation, and DE in general, affects the growth of matter density perturbation $\delta_m\equiv \frac{\delta \rho_m}{\bar{\rho}_m}$, is to rewrite the fluid equations for matter and DE as a second order differential equation for $\delta_m$. To do so, we assume homogeneity, isotropy and neglect neutrinos, which is a viable approximation since in such small scales our data is not  affected by them. Then, the growth of matter can be followed with the second order differential equation \cite{Amendola:2007rr,Tsujikawa:2010zza,Nesseris:2008mq,Nesseris:2009jf}

\begin{equation}
  \delta^{\prime \prime}_m(a)+\left[\frac{3}{a}+\frac{H^{\prime}(a)}{H(a)}\right]\!\delta_m^{\prime}(a)- \frac{3\Omega_{m_{0}}H_{0}^{2}\,G_{\mathrm{eff}}(a)}{2a^{5} H(a)^{2}\,G_N} \delta_m(a)=0,\label{eq:growth}
\end{equation}
where the effects of DE or a modified gravity theory, such as $f(R)$, at the perturbations level can be taken into account by the effective Newtonian constant $G_{\textrm{eff}}(a)$. 

To find the effects of the nonadiabatic pressure perturbation we follow  Ref.~\cite{Sapone:2009mb}, where it was shown that for a DE fluid with constant equation of state $w$ during matter domination $Q \equiv G_{\mathrm{eff}}(a) / G_{N}$ is given by
\be
Q-1=\left(\frac{1-\Omega_{m}}{\Omega_{m}}\right)\left(\frac{1+w}{1-3 w}\right) a^{-3 w} \equiv Q_{0} a^{-3 w}.
\ee
To find a similar expression of $Q$ during dark energy domination, which is a solution on small scales $k \gg aH/\hat{c}_s$, that takes into account the nonadiabatic component $\hat{\delta P}_{\textrm{nad}}/\bar{\rho}$ we do the following. Defining the scalar velocity perturbation as $V \equiv i k_{j} T_{0}^{j} / \rho=(1+w) \theta$, Eqs.~(\ref{eq:eqd2}) and (\ref{eq:eqth2}) can be rewritten, in the conformal Newtonian gauge, as
\ba
\delta_\textrm{DE}'&=&-\frac{V_\textrm{DE}}{Ha^2}\left(1+\frac{9a^2H^2\left(\hat{c}_s^2-c_a^2\right)}{k^2}\right)-\frac{3}{a}\left(\hat{c}_s^2-w\right)\delta_\textrm{DE} \nn \\
&+&3(1+w)\phi'-\frac{3}{a}\frac{\hat{\delta P}_{\textrm{nad}}}{\bar{\rho}}, \label{eq:eqd21}\\
V_\textrm{DE}'&=&-(1-3\hat{c}_s^2)\frac{V_\textrm{DE}}{a}+\frac{k^2}{Ha^2}\hat{c}_s^2\delta_\textrm{DE}+(1+w)\frac{k^2}{Ha^2}\psi \nn\\
&+&\frac{\hat{\delta P}_{\textrm{nad}}}{\bar{\rho}}\frac{k^2}{Ha^2}-\frac{(1+w)k^2}{Ha^2}\sigma,\label{eq:eqth21}
\ea
where the prime $'$ is the derivative with respect to the scale factor $a$ and we are assuming there is no DE anisotropic stress, i.e $\sigma=0$, and hence $\phi=\psi$. In Eq.~(\ref{eq:eqth21}), in order to not have large velocity perturbations it is expected that the terms that scale as $k^2$ cancel out, hence
\be
\delta_\textrm{DE}=\frac{3}{2}(1+w)\frac{H_0^2\Omega_m}{\hat{c}_s^2k^2}\delta_0-\frac{\hat{\delta P}_{\textrm{nad}}/\bar{\rho}}{\hat{c}_s^2}\label{eq:deltaini},
\ee
where we have used that $k^2\phi=-\frac{3}{2}\delta_0H_0^2\Omega_m$ which is the solution for the perturbation equations in matter domination \cite{Sapone:2009mb}. Then using Eqs.~(\ref{eq:eqd21}) and (\ref{eq:deltaini}) we find
\be
V_\textrm{DE}=-3Ha(\hat{c}_s^2-w)\delta-3Ha \frac{\hat{\delta P}_{\textrm{nad}}}{\bar{\rho}}.
\ee
Now we can compute $Q$ in the dark energy domination regime as
\be
Q-1=\frac{\rho_{\textrm{DE}} \Delta_{\textrm{DE}}}{\rho_{m} \Delta_{m}}
\ee
where $\Delta\equiv \delta+\frac{3aHV}{k^2} $ is the gauge invariant density perturbation. In matter domination we have that  $\Delta_{m}=\delta_0 a$, while for DE we have that 
\be
\Delta_{\textrm{DE}} \simeq \frac{3}{2}(1+w)\frac{H_0^2\Omega_m}{\hat{c}_s^2k^2}\delta_0-\frac{\hat{\delta P}_{\textrm{nad}}/\bar{\rho}}{\hat{c}_s^2}\label{eq:Deltade},
\ee
which is similar to the initial condition given by Eq.~\eqref{eq:dDEICMD}. From Eq.~(\ref{eq:Deltade}) we see that the dominant term comes from the contribution of the nonadiabatic part, as the latter scales as $k^2$, see Eq.~(\ref{eq:Pansatz}), hence $Q$ can be expressed as
\be\label{eq:Qnonad}
Q-1\simeq -\frac{1-\Omega_m}{\Omega_m}\frac{\hat{\delta P}_{\textrm{nad}}/\bar{\rho}}{\delta_0\hat{c}_s^2}a^{-1-3w}.
\ee

\subsubsection{Analytic solutions for the growth}
Modeling the nonadiabatic pressure perturbation as in Eq.~(\ref{eq:Pansatz}), $Q$ can be written as
\be
Q(k,a) =1-\frac{1-\Omega_m}{\Omega_m}\frac{c_0 k^2}{\delta_0\hat{c}_s^2 H_0^2}a^{n-1-3w}.\label{eq:Qnonad1}
\ee
In order to solve Eq.~\eqref{eq:growth} with $Q \equiv G_{\mathrm{eff}}(k,a) / G_{N}$ given by Eq.~\eqref{eq:Qnonad1} we need to make an approximation due to the appearance of the term $a^{n-1}$, which makes it difficult to find analytic solutions. As we expect that $n\sim O(1)$ at late times (see Fig.~\ref{fig:fRdesigner}), then we make a series expansion of the term $a^{n-1}$ around $n=1$ of the form 
\ba
a^{n-1} &\simeq& 1 + (n - 1)\ln{a}+\cdots \nn \\
&\simeq& 1 - (n - 1)\ln{(1+z)}+\cdots, \label{eq:napp}
\ea
where in the second step we used that $a=\frac1{1+z}$. Since we are interested in the evolution of the growth at low redshifts, we replace the term $\ln{(1+z)}$ with an average $b_0=\langle \ln{(1+z)} \rangle$, which in the range $z\in [0,2]$ is approximately $b_0\simeq 0.6479$. Hence, under this approximation Newton's constant becomes
\be
Q(k,a) \simeq 1-\frac{1-\Omega_m}{\Omega_m}\frac{c_0 k^2}{\delta_0\hat{c}_s^2 H_0^2}\left(1 - b_0\;(n - 1)\right)a^{-3w}.\label{eq:Qnonad2}
\ee
Then, by making the change of variables $a^{-3w}\equiv x$ and inserting Eq.~(\ref{eq:Qnonad}) into Eq.~(\ref{eq:growth}) we find
\ba\label{eq:growth_n}
 \delta_m(a) &=&a _{2} F_{1}\left[\frac{1}{4}-\frac{5}{12 w}+B, \frac{1}{4}-\frac{5}{12 w}-B, 1-\frac{5}{6 w};\right.\nn\\ &&\left.-\frac{1-\Omega_{\mathrm{m} 0}}{\Omega_{\mathrm{m}_{0}}} a^{-3 w}\right], \label{eq:approx}
\ea
where
\ba
B&=&-\frac{1}{12 w} \sqrt{(1-3 w)^{2}+24\delta B},\\
\delta B&=&-\frac{c_0 k^2}{\delta_0\hat{c}_s^2 H_0^2}\left(1 - b_0\;(n - 1)\right).\label{eq:deltaB}
\ea

To compare our analytical results with the full numerical solution from the evolution equations \eqref{eq:eqd2}-\eqref{eq:eqth2} in the next sections we will use the combination $f\sigma_8(a)$ which is a measurable quantity and is defined as
\ba
f \sigma_{8}(a) & \equiv& f(a) \cdot \sigma(a) \nn\\
&=&\frac{\sigma_{8,0}}{\delta_{m}(1)} a \delta_{m}^{\prime}(a),\label{eq:fs8def}
\ea
where $\sigma(a)=\sigma_{8,0} \frac{\delta_{m}(a)}{\delta_{m}(1)}$ is the redshift-dependent rms fluctuations of the linear density field at $R=8h^{-1}$Mpc while the parameter $\sigma_{8,0}$ is its value today. Since in order to derive the solution of Eq.~\eqref{eq:approx} we have neglected radiation, neutrinos and baryons, we note that the solution is only valid at late times.

\subsubsection{The growth rate index $\gamma$ \label{sec:growthgamma}}
Finally, we briefly discuss the growth index $\gamma$ in the presence of DE perturbations. The latter affect the evolution of the matter density contrast $\delta_m\equiv\frac{\delta \rho_m}{\rho_m}$ and its growth rate $f(a)\equiv \frac{d\ln \delta_m}{d\ln a}$. When we ignore DE perturbations, the latter can be approximated as \cite{Wang:1998gt, Belloso:2011ms,Nesseris:2015fqa}
\be
f(a) =\Omega_{\mathrm{m}}(a)^{\gamma(a)},\label{eq:gammadef}
\ee
where the growth index $\gamma$ is given by
\ba
\gamma(a) &=& \gamma_m(a) \nn \\
&=& \frac{\ln f(a)}{\ln \Omega_{\mathrm{m}}(a)} \nn \\
&\simeq& \frac{3(1-w)}{5-6 w}+\cdots,
\ea
which for \lcdm reduces to $\gamma\sim\frac{6}{11}$ and by $\gamma_m$ we denote the contribution to the growth index coming from the CDM and the background evolution only. When we include the DE perturbations assuming they are sourced from an anisotropic stress, the growth index picks up a correction \cite{Nesseris:2015fqa}
\be
\gamma=\gamma_{m}+\gamma_{\textrm{DE}},
\ee
where the contribution coming from the DE perturbations is given by
\be
\gamma_{\textrm{DE}}\simeq-\frac{3(1+w)}{18w^2-21w+5}+\cdots.
\ee
From now on we will refer to $\Omega_m(a)$ as $\Omega$ as a shorthand. If we include DE perturbations the growth index for the matter can be written to first order as
\ba
\gamma&=&\frac{\ln (f(\Omega))}{\ln (\Omega)}\nn \\
&=&\frac{3(\delta B+w-1)}{6 w-5}- \nn\\
&&\frac{3(\Omega-1)((\delta B+w-1)(9 \delta B(4 w-3)-3 w+2))}{2\left((5-6 w)^{2}(12 w-5)\right)}+\cdots, \nn \\\label{eq:gammatot}
\ea
We can split the growth index into two parts: the contribution from the CDM component and the background expansion denoted as $\gamma_m$, and the contribution from the nonadiabatic component, denoted as $\gamma_{\textrm{DE}}$. Then we have
\be
\gamma=\gamma_{m}+\gamma_{\textrm{DE}},
\ee
and we find from Eq.~\eqref{eq:gammatot} that
\ba
\gamma_{m}&=&\frac{3(w-1)}{6 w-5}+\frac{3(3 w-2)(w-1)(\Omega-1)}{2(5-6 w)^{2}(12 w-5)}+\cdots, \\
\gamma_{D E}&=&\frac{3 \delta B}{6 w-5}+(\Omega-1)\left(-\frac{3 \delta B(6 w(6 w-11)+29)}{2\left((5-6 w)^{2}(12 w-5)\right)}-\right. \nn\\
&&\left.\frac{27 \delta B^{2}(4 w-3)}{2\left((5-6 w)^{2}(12 w-5)\right)}\right)+\cdots,
\ea
where $\delta B$ is given by Eq.~\eqref{eq:deltaB}. These expressions are similar to those when DE perturbations are included, originally derived in Ref.~\cite{Nesseris:2015fqa}, but now  the extra contribution comes instead from the nonadiabatic pressure perturbation. 

\subsection{Discussion on the scale-dependent growth}
Large scale structure surveys measure the growth rate $f\sigma_8(z)$ by using the values from the multipoles of the redshift-space galaxy two-point correlation function at late times, see for example Ref.~\cite{Jelic-Cizmek:2020pkh}. This requires modeling the multipoles at the redshift $z$ either by assuming a fiducial cosmological model, so as to compute the shape of the real-space power spectrum, or by assuming the growth is scale-independent from the onset of some early redshift and then assuming another fiducial model at early times. Both approaches can in principle be problematic if dark energy causes the growth to be strongly scale-dependent, so it would be ideal to directly model the multipoles in the particular scale-dependent cosmology at hand and do the parameter inference at the multipole level. However, this approach is computationally extremely prohibitive for doing Monte Carlo analyses, so we do not consider it here   opting instead to examine how the scale-dependence of the growth affects our results.

One way to examine this dependence would be to create mock $f\sigma_8(z)$ data using an N-body simulation of the nonadiabatic dark energy model and then check whether the input fiducial cosmology can be recovered, as was done for example in Ref.~\cite{Bose:2017myh}. In particular, the authors of Ref.~\cite{Bose:2017myh} found that this scale-dependence of the growth can significantly bias the parameter constraints. However, such an analysis is beyond the scope of our paper, so instead in the next section we will extensively study how strong this scale-dependence is by comparing the analytical and Mathematica numerical solutions to those of CLASS, as well as by studying the scale-dependence of the growth as a function of the wave-number $k$.

\section{Comparison with CLASS and numerical solutions \label{sec:class}}

Here we present in detail how the nonadiabatic DE pressure perturbation, given by the ansatz of Eq.~\eqref{eq:Pansatz}, affects several key cosmological quantities, such as the scale-dependent growth $f\sigma_8(k,z)$, the matter power spectrum $P(k,z)$ and the CMB TT power spectrum $C_\ell^\textrm{TT}$. 

\begin{figure*}[!t]
\centering
\includegraphics[width = 0.49\textwidth]{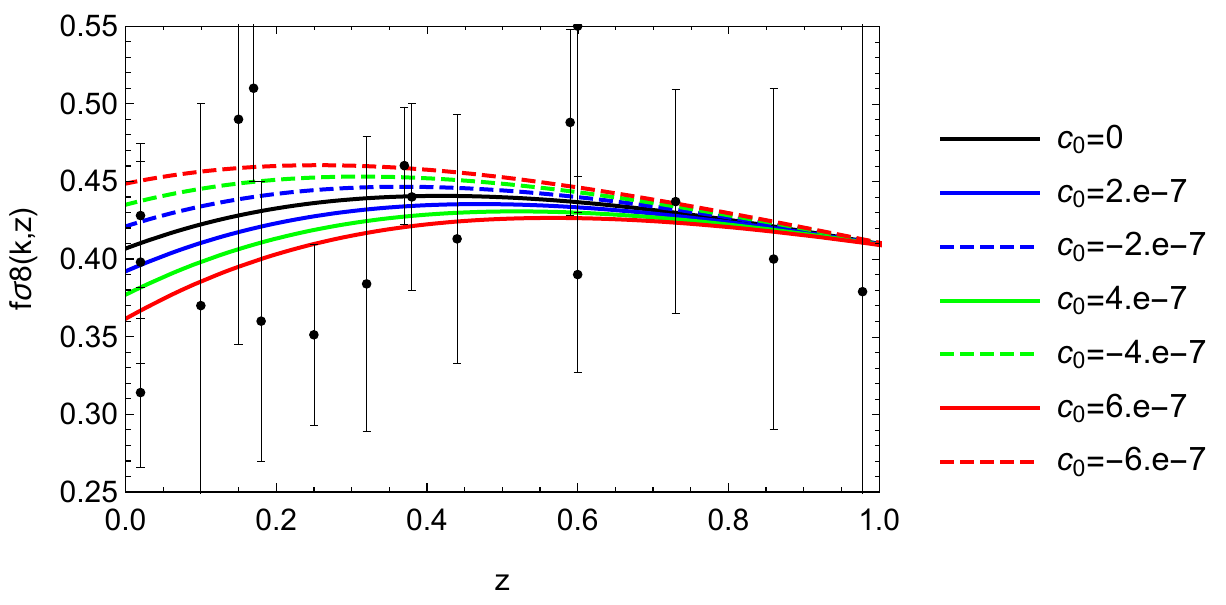}
\includegraphics[width = 0.48\textwidth]{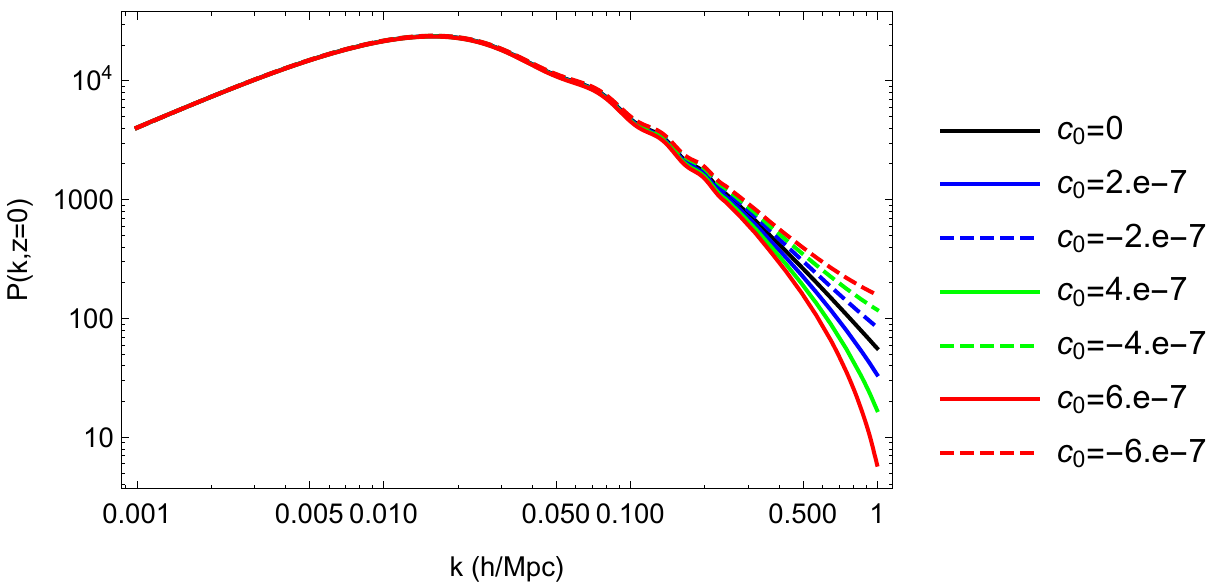}
\caption{Left panel: the evolution of the scale-dependent growth rate $f\sigma_8(k,z)$ for various values of $c_0$ and $n=0.5$. In this case the growth was calculated with CLASS via $\delta(k,z)\equiv \sqrt{\frac{P(k,z)}{P(k,0)}}$ for $k=0.1 h/\textrm{Mpc}$. The points correspond to the  ``Gold 2018" growth rate $f\sigma_8$ compilation shown in Table~\ref{tab:fs8tab}. Right panel: the matter power spectrum $P(k,z)$ at $z=0$, for various values of $c_0$ and $n=0.5$. In both cases we assumed $\Omega_{m0}=0.3$, $w=-0.8$, $h=0.67$.  \label{fig:fs8pk}}
\end{figure*}

\begin{figure*}[!t]
\centering
\includegraphics[width = 0.48\textwidth]{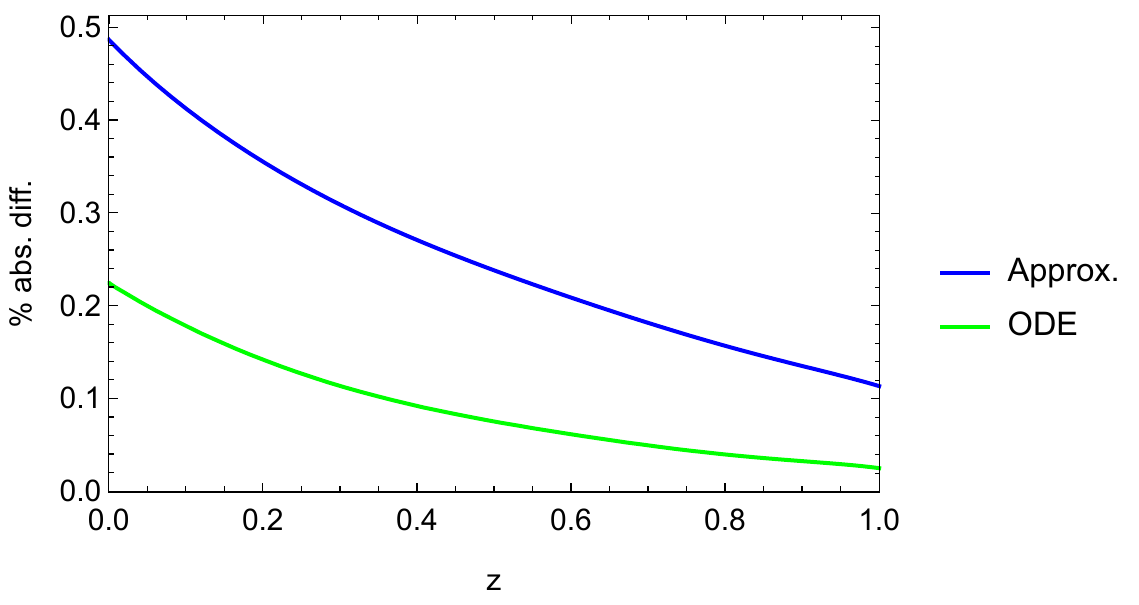}
\includegraphics[width = 0.48\textwidth]{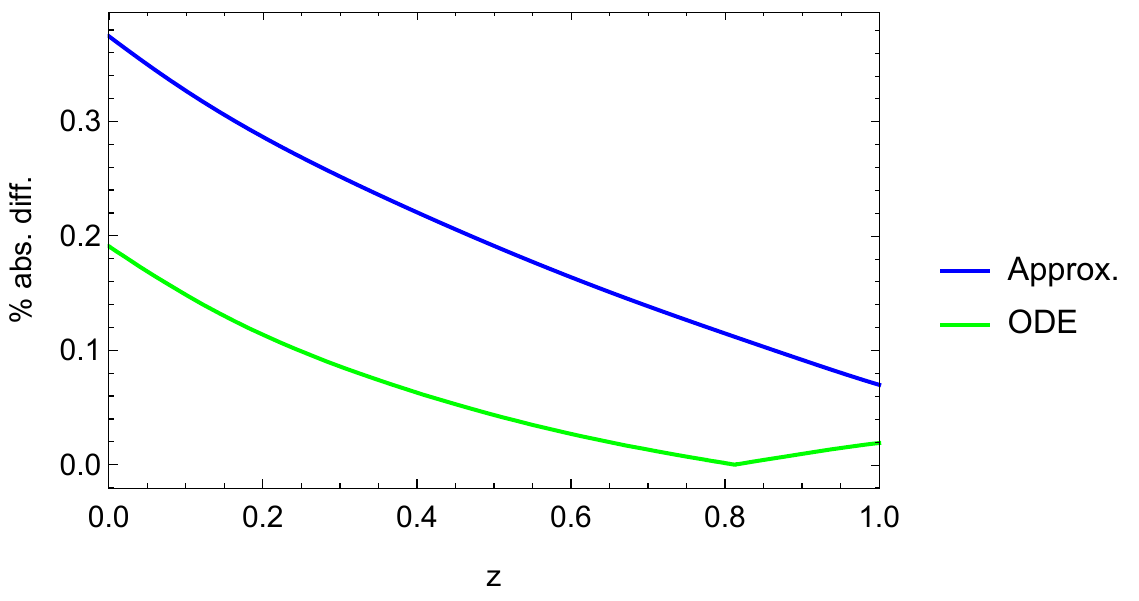}
\caption{Left panel: Tthe absolute percentage difference of $f\sigma_8(z)$ between the numerical solution from the evolution equations Eq.~\eqref{eq:eqth2} (green line, denoted ``ODE") and the analytical approximation of Eq.~\eqref{eq:approx} (denoted ``Approx.'') with respect to the numerical solution from our CLASS implementation for $c_0=2\cdot10^{-7}$. Right panel: same as the left panel, but for  $c_0=-2\cdot10^{-7}$. For both plots we assume $\Omega_m=0.3$, $w=-0.8$, $\hat{c}_s^2=1$, $h=0.67$, $k=0.1 h/\textrm{Mpc}$ and $\sigma_8=0.8$.
\label{fig:diffs}}
\end{figure*}

To do this, we implemented the nonadiabatic pressure perturbation as given by Eq.~\eqref{eq:Pansatz}, along with the initial conditions in radiation domination in the synchronous gauge, given by Eqs.~\eqref{eq:ICs1}-\eqref{eq:ICs2}, in the Boltzmann code CLASS \cite{Blas:2011rf, Lesgourgues:2011re}. To test our modifications, we also compare the numerical results from CLASS with the numerical solution in Mathematica of the fluid equations \eqref{eq:eqd2} and \eqref{eq:eqth2}, but also with the analytical solutions of Sec.~\ref{sec:analytic}. 

We should note that there is a difference between the normalization used in CLASS, which uses units of Mpc and thus affects the initial values of the perturbations $\delta_0$, and in the numerical solution  of the evolution equations \eqref{eq:eqd2} and \eqref{eq:eqth2} in Mathematica, where we set $\delta_0=1$, so that $\delta_m(a)\sim a$ in matter domination, and $k$ is expressed in units of $H_0$. For example, a wavenumber of $k=0.1\textrm{Mpc}^{-1}$ in CLASS corresponds to $k=0.1 \frac{3000}{h}H_0=\frac{300}{h} H_0$ in our notation\footnote{Similarly, a wavenumber of $k=0.1 h\textrm{Mpc}^{-1}$ is equivalent to $k=0.1\cdot3000 H_0=300 H_0$.}. Then, the coefficient $c_0$ is rescaled by a factor of $c_{0,\textrm{CLASS}}\rightarrow c_{0,\textrm{Math}} \left(\frac{3000}{h}\right)^2 \frac{\delta_{0,\textrm{CLASS}}}{\delta_{0,\textrm{Math}}}$ between the two frameworks. In what follows, we will express all values of $c_0$ in the dimensionless picture, i.e. $c_0=c_{0,\textrm{Math}}$, as that is easier to test numerically with any ordinary differential equation solver, not only CLASS. In what follows we will assume a fiducial cosmology with $\Omega_m=0.3$, $w=-0.8$, $\hat{c}_s^2=1$, $h=0.67$, $k=0.1 h/\textrm{Mpc}$ and $\sigma_8=0.8$, unless otherwise specified.

First, in Fig.~\ref{fig:fs8pk} we show the dependence of the growth rate $f\sigma_8(k,z)$ and the matter power spectrum $P(k,z)$ on the parameter $c_0$ keeping $n$ fixed. In the left panel we show the evolution of the scale-dependent growth rate $f\sigma_8(k,z)$ for various values of $c_0$ and $n=0.5$. In this case the growth was calculated with CLASS via $\delta(k,z)\equiv \sqrt{\frac{P(k,z)}{P(k,0)}}$ for $k=0.1 h/\textrm{Mpc}$. As can be seen, the amplitude of the pressure perturbation ansatz $c_0$ has a strong effect on the growth rate $f\sigma_8(k,z)$ at late times $z<1$, and thus we expect it to be tightly constrained in the MCMC analysis in the next section. 

We also tested our codes by calculating the growth for $w=-1$ and as expected, we find that the main effect indeed comes from the nonadiabatic perturbations. This test is important as the usual dark energy perturbations within GR exactly vanish for the cosmological constant model $(w=-1)$, so any difference of the growth from its expected \lcdm value would be a smoking gun signal for modified gravity.

On the other hand, in the right panel of Fig.~\ref{fig:fs8pk} we show the matter power spectrum $P(k,z)$ at $z=0$, for various values of $c_0$ and $n=0.5$. As can be seen, the effect of the nonadiabatic perturbations in this case is to suppress or enhance power, depending on the sign of $c_0$, an effect similar to that observed in Ref.~\cite{Kunz:2015oqa} for a mixed DE-DM model and in Ref.~\cite{Brown:2011dn} for a similar ansatz. Note that in general the matter power spectrum $P(k,z)$ at scales $k\sim 0.1-10\;h/\textrm{Mpc}$ can be constrained by Lyman alpha data \cite{Zaroubi:2005xx}, however as those observations are beyond the scope of this work we do not consider them in this analysis. 

Next, we compare the results for the growth rate between CLASS, Mathematica and the analytical approximation to the growth equation. In Fig.~\ref{fig:diffs} we show the absolute percentage difference of $f\sigma_8(z)$ between the numerical solution from the evolution equations  \eqref{eq:eqth2} (green line, denoted ``ODE") and the analytical approximation of Eq.~\eqref{eq:approx} (denoted ``Approx.'') with respect to the numerical solution from CLASS for $c_0=2\cdot10^{-7}$. In the right panel we show the same functions as in the left one, but for  $c_0=-2\cdot10^{-7}$. As seen in Fig.~\ref{fig:diffs}, with the approximation we have sub-percent agreement between the analytic approximation and the numerical one at late times. Note however, that neither the analytical solutions in Mathematica (denoted ``ODE") nor the analytical solutions of Eq.~\eqref{eq:approx} (denoted ``Approx.'') include radiation, neutrinos or baryons, and hence their range of validity in terms of the wavenumber $k$ is limited to $k\ge k_{eq}$. Here we only consider them in order to gain physical insight on the behavior of this model.

\begin{figure*}[!t]
\centering
\includegraphics[width = 0.46\textwidth]{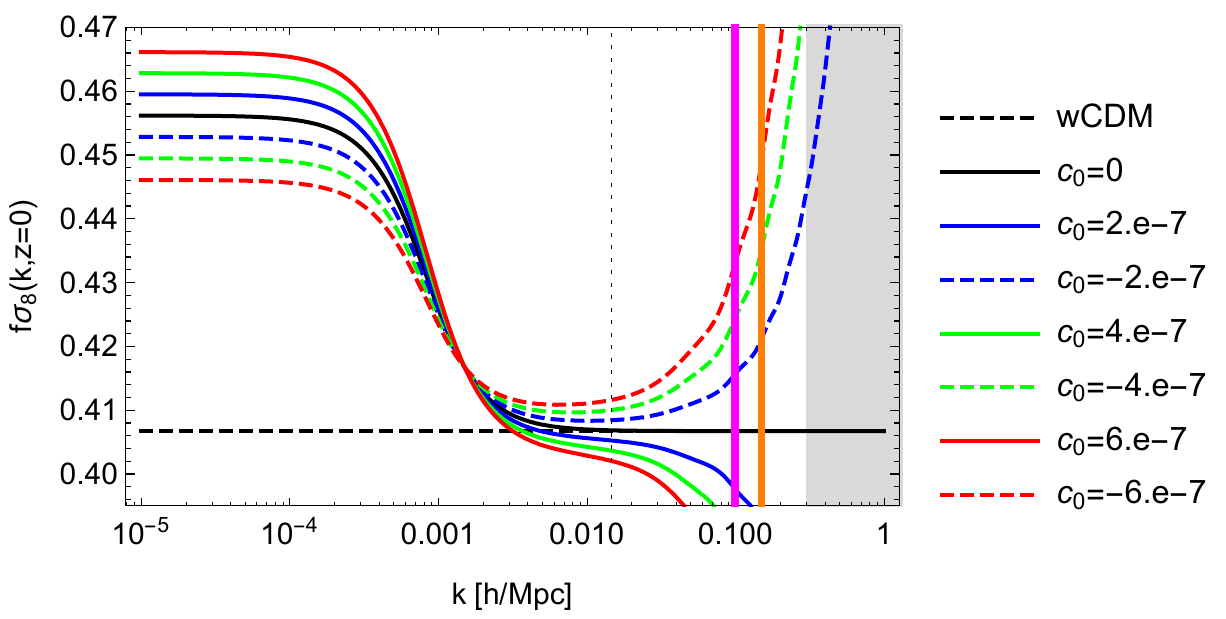}
\includegraphics[width = 0.53\textwidth]{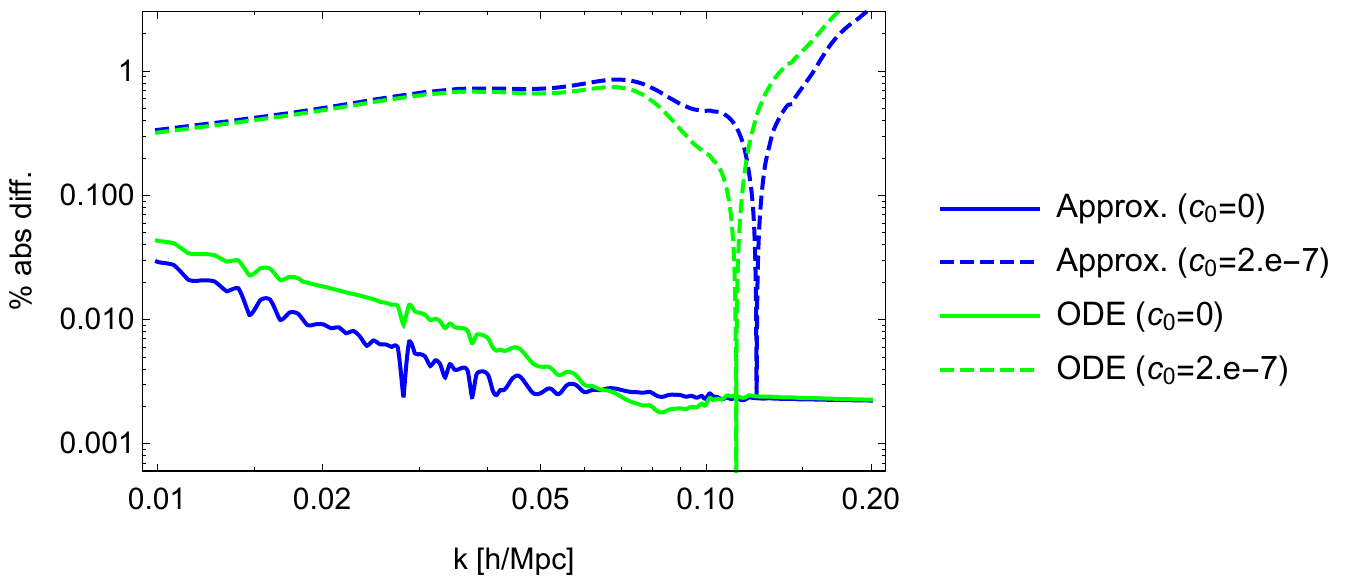}
\caption{Left: The present value of the scale-dependent growth $f\sigma_8(k,z=0)$ as a function of the wavenumber $k$. The dashed black line is the scale independent growth in GR (neglecting radiation, neutrinos and baryons) given by the solution to Eq.~\eqref{eq:eqth}, the vertical dotted line corresponds to the scale of equality $k_\textrm{eq}\simeq 0.073\;\Omega_{m,0} h^2/\textrm{Mpc}$, while the grey region denotes the non-linear regime \cite{Takahashi:2012em}. The vertical magenta and orange lines correspond to the effective wavenumber for SDSS and WiggleZ, while the colored lines correspond to various values of $c_0$. Right: The scale-dependence of the absolute percent difference of the approximate solution for $c_0=0$ (blue line) and $c_0=2\cdot 10^{-7}$ (blue dashed line) and the Mathematica numerical solution for $c_0=0$ (green line) and $c_0=2\cdot 10^{-7}$ (green dashed line) against the solution from CLASS. In both cases we assumed $\Omega_{m0}=0.3$, $w=-0.8$, $h=0.67$.\label{fig:fs8kdep}}
\end{figure*}

In Fig.~\ref{fig:fs8kdep} we show the scale-dependence of the growth rate and a comparison with the solution from CLASS. In particular, in the left panel we show the present value of the scale-dependent growth $f\sigma_8(k,z=0)$ as a function of the wavenumber $k$. The dashed black line is the scale independent growth in GR (neglecting radiation, neutrinos and baryons) given by the solution to Eq.~\eqref{eq:eqth}, the vertical dotted line corresponds to the scale of equality $k_\textrm{eq}\simeq 0.073\;\Omega_{m,0} h^2/\textrm{Mpc}$, while the grey region denotes the non-linear regime \cite{Takahashi:2012em}. The vertical magenta and orange lines correspond to the effective wavenumber for SDSS and WiggleZ of $k=0.1 h/\textrm{Mpc}$ and $k=0.15 h/\textrm{Mpc}$ respectively, while the colored lines correspond to various values of $c_0$. At $k\sim k_{eq}$ (vertical dotted line) the dashed black and solid black lines agree perfectly, while the deviations at small $k$ are due to radiation, neutrinos etc included in CLASS.

In the right panel of Fig.~\ref{fig:fs8kdep} we show the scale-dependence of the absolute percent difference of the approximate solution for $c_0=0$ (blue line) and $c_0=2\cdot 10^{-7}$ (blue dashed line) and the Mathematica numerical solution for $c_0=0$ (green line) and $c_0=2\cdot 10^{-7}$ (green dashed line) against the solution from CLASS. As can be seen, especially close to the
scales where the data are, i.e. $0.1<k\; (\textrm{Mpc}/h) <0.15$, the agreement is below $1\%$. At higher $k$ the difference rises somewhat above $1\%$, but we then quickly enter the non-linear regime where, as mentioned earlier, our calculations are not valid.

\begin{figure}[!t]
\centering
\includegraphics[width=0.49\textwidth]{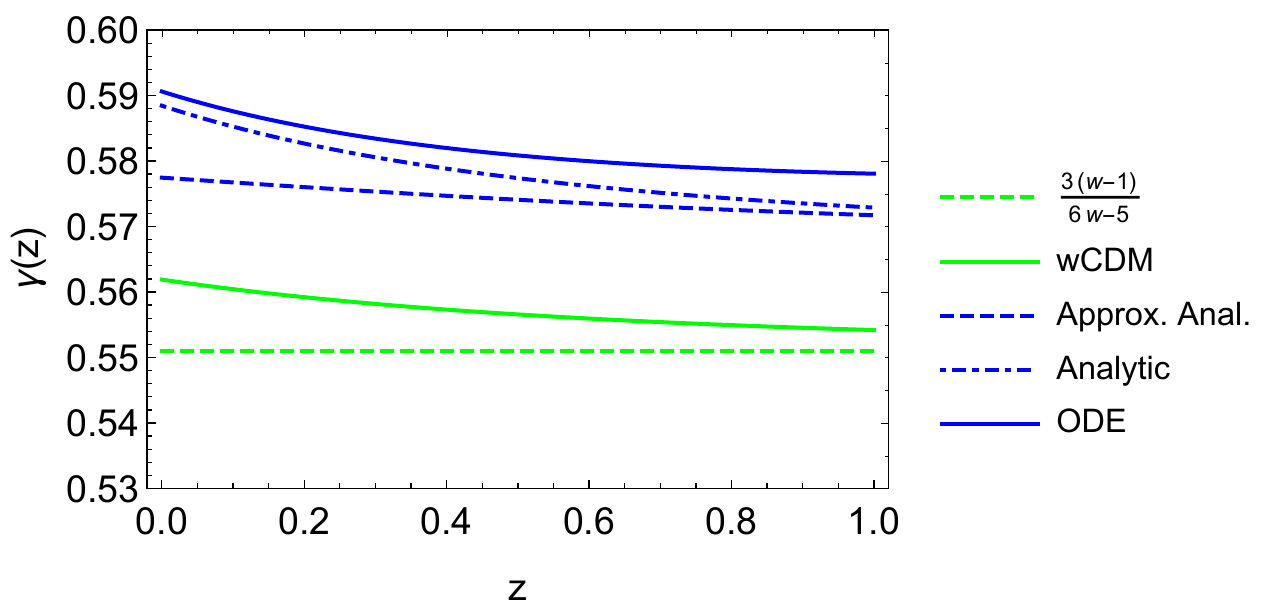}
\caption{The growth index $\gamma$ as a function of redshift for $\Omega_{m0}=0.3$, $k=0.1\;h/ \textrm{Mpc}$, $c_0=2\cdot 10^{-7}$, $n=0.5$ and $w=-0.8$ for five different cases: $\gamma=\frac{3(w-1)}{6w-5}$ (dashed green line), the analytical expression when inverting Eq.~\eqref{eq:gammadef} for the $w$CDM model (solid green line), the analytical approximation to first order of Eq.~\eqref{eq:gammatot} (dashed blue line), the analytical expression when inverting Eq.~\eqref{eq:gammadef} with the growth given by Eq.~\eqref{eq:approx} for the nonadiabatic model model (dot-dashed blue line) and the numerical solution of the fluid equation for the nonadiabatic model model (solid blue line).
\label{fig:gamma}}
\end{figure}

In Fig.~\ref{fig:gamma} we also compare the predictions for the growth index $\gamma$ as a function of redshift for  $c_0=2\cdot 10^{-7}$ and $n=0.5$ for five different cases: $\gamma=\frac{3(w-1)}{6w-5}$ (dashed green line), the analytical expression when inverting Eq.~\eqref{eq:gammadef} for the $w$CDM model (solid green line), the analytical approximation to first order of Eq.~\eqref{eq:gammatot} (dashed blue line), the analytical expression when inverting Eq.~\eqref{eq:gammadef} with the growth given by Eq.~\eqref{eq:approx} for the nonadiabatic model model (dot-dashed blue line) and the numerical solution of the fluid equation for the nonadiabatic model model (solid blue line). As can be seen, in all cases the agreement between the exact numerical result (solid blue line) and the two approximations is on average of the order of a percent.

\begin{figure*}[!t]
\centering
\includegraphics[width = 0.48\textwidth]{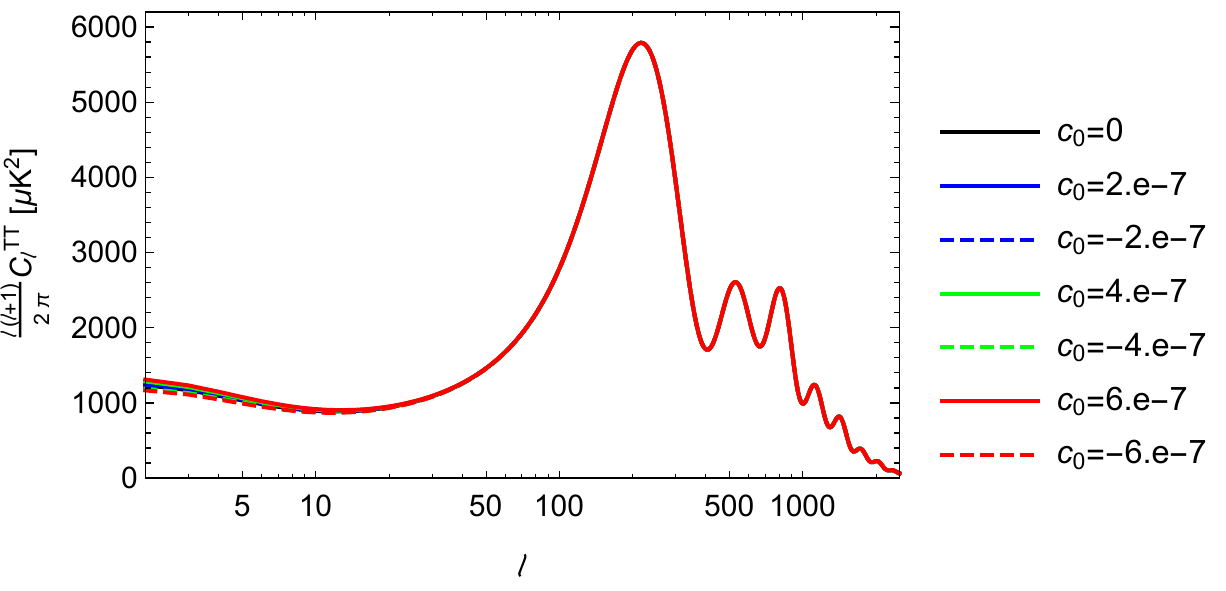}
\includegraphics[width = 0.48\textwidth]{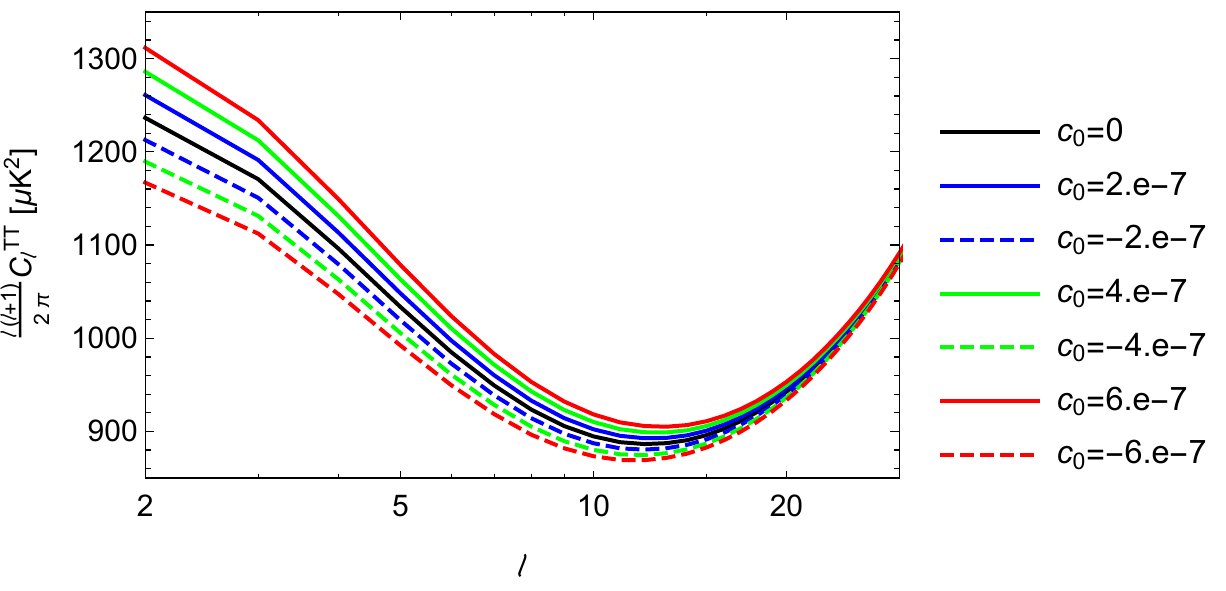}
\caption{The effect of the nonadiabatic pressure perturbation, given by Eq.~\eqref{eq:Pansatz}, on the TT CMB spectrum (left) and its low multipoles (right). As can be seen, the effect is either to enhance or suppress power on small multipoles, with the rest of the TT spectrum remaining unchanged. In both cases we assumed $\Omega_{m0}=0.3$, $w=-0.8$, $h=0.67$.\label{fig:cls}}
\end{figure*}

Finally, in Fig.~\ref{fig:cls} we show the effect of the nonadiabatic pressure perturbation, given by Eq.~\eqref{eq:Pansatz}, on the TT CMB spectrum (left) and its low multipoles (right). Overall, the effect is either to enhance or suppress power on large scales, i.e. small multipoles, with the rest of the TT spectrum remaining unchanged. Thus, in our MCMC analysis in the next section, we expect the main constraint from the Planck 18 data to come from the integrate Sachs-Wolfe (ISW) part of the TT CMB spectrum. 

\section{MCMC results \label{sec:mcmc}}
Here we discuss how MontePython \cite{Audren:2012wb, Brinckmann:2018cvx} was used to place constraints via an MCMC approach on the parameters of the ansatz for the nonadiabatic pressure perturbation given by Eq.~\eqref{eq:Pansatz}. We used the Planck 2018 CMB data and in particular we add the ``Planck\_highl\_TTTEEE", ``Planck\_lowl\_EE", ``Planck\_lowl\_TT" temperature and polarization likelihoods (collectively called CMB later on), as well as the CMB lensing ``Planck\_lensing" likelihood \cite{Aghanim:2018eyx}.

We also add the BOSS DR-12 data \cite{Alam:2016hwk}, the 6dF BAO points from Ref.~\cite{Beutler:2011hx}, the MGS BAO points from \cite{Ross:2014qpa} and the Pantheon SnIa compilation of Ref.~\cite{Scolnic:2017caz}. Finally, we also include an RSD likelihood for MontePython, presented here for the first time, based on the ``Gold 2018" growth rate $f\sigma_8$ compilation given in Table I of Ref.~\cite{Sagredo:2018ahx}. We discuss the new likelihood in detail in the Appendix.

As seen in the previous section, the nonadiabatic DE pressure perturbation, as given by Eq.~\eqref{eq:Pansatz}, may affect the clustering of objects, by either suppressing or enhancing it. The non-linear regime in Boltzmann codes like CLASS is taken into account with routines like Halofit \cite{Takahashi:2012em}, which is calibrated with $\Lambda$CDM N-body simulations in order to emulate the effects of non-linearities on small scales [$0.1 < k\, (\text{Mpc}/h) \lesssim 10$] for a range of $\Lambda$CDM parameters. Halofit should not be expected to work well, if at all, in models that deviate significantly from this scenario. Indeed, simulations of Warm Dark Matter (WDM) models demonstrate that Halofit overestimates the power spectrum at small scales \cite{Viel:2011bk}. Hence, one should be careful when using Halofit, especially when the model under consideration induces scale-dependent corrections to the matter power spectrum, as in our case here. As a result, we have decided to turn off Halofit in our analysis.

We then ran MCMC chains for the $w$CDM model and the nonadiabatic model for two data combinations each: CMB+BAO+SnIa and CMB+BAO+SnIa+RSD in order to assess the constraining power of the new RSD likelihood. For the $w$CDM model we ran four chains with roughly 200,000 points, while for the nonadiabatic model we ran 19 chains with roughly 2,000,000 points in total in order to make sure all the parameters, described below, were well converged.

Specifically, for the MCMC runs of the $w$CDM model we consider the following parameters: the DE equation of state parameter $w$, assuming it is constant, the baryon and cold dark matter density parameters $\Omega_{b,0}h^2$ and $\Omega_{c,0}h^2$ respectively, the angular scale of the acoustic oscillations $\theta$, the
optical depth to Thomson scattering from reionization $\tau$ and the two parameters of the primordial power spectrum $A_s$ and $n_s$. In a nutshell, our parameter vector for the $w$CDM model is then $p_{w\textrm{CDM}}=\left(w,\Omega_{b,0}h^2,\Omega_{c,0}h^2,\theta,A_s, n_s\right)$. On the other hand, for the MCMC runs of the nonadiabatic model, we include the parameters of the  $w$CDM model, along with the two nonadiabatic parameters $c_0$ and $n$ of Eq.~\eqref{eq:Pansatz}. Then, our parameter vector for the nonadiabatic  model is $p_{\textrm{non-ad}}=\left(w,\Omega_{b,0}h^2,\Omega_{c,0}h^2,\theta,A_s, n_s,c_0,n\right)$.

\begin{figure*}[!t]
\centering
\includegraphics[width=0.85\textwidth]{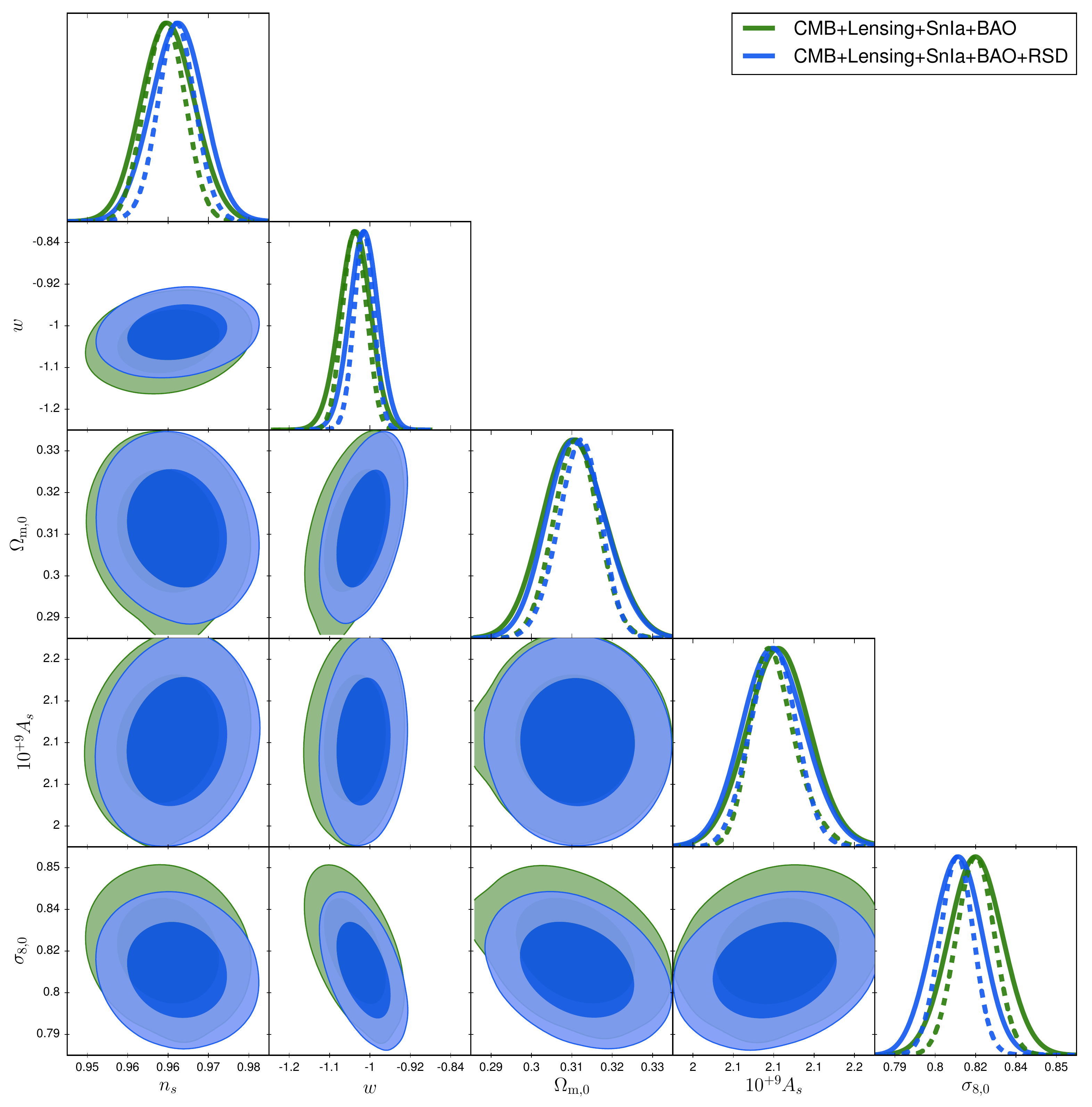}
\caption{The confidence contours for the $w$CDM model using the data combinations of Planck+Lensing+BAO+SnIa (green contours) and  Planck+Lensing+BAO+SnIa+RSD (blue contours).\label{fig:mcmcwcdm}}
\end{figure*}

In Fig.~\ref{fig:mcmcwcdm} we show the confidence contours for $w$CDM using CMB+Lensing+BAO+SnIa (green contours) and the CMB+Lensing+BAO+SnIa+RSD (blue contours), while in Tables~\ref{tab:wcdm1}-\ref{tab:wcdm2} we present the $68\%$ mean values and $95\%$ confidence regions, for some of the parameters of the model. As can be seen, the contours are a bit shifted to higher values of $\sigma_{8,0}$ and $w$ when the RSD data included. This is consistent with the well-known tension for $\sigma_8$ between low and high redshift probes \cite{Nesseris:2017vor}.

\begin{table}[!b]
\begin{tabular}{|l|c|c|c|c|} 
 \hline 
Param & best-fit & mean$\pm\sigma$ & 95\% lower & 95\% upper \\ \hline 
$n_s$ &$0.9622$ & $0.965_{-0.0041}^{+0.0039}$ & $0.9571$ & $0.973$ \\ $w$ &$-1.024$ & $-1.03_{-0.032}^{+0.033}$ & $-1.095$ & $-0.9673$ \\ 
$\Omega_\mathrm{m,0}$ &$0.3109$ & $0.3058_{-0.0082}^{+0.0076}$ & $0.2903$ & $0.3215$ \\ 
$10^{+9} A_s$ &$2.086$ & $2.104_{-0.033}^{+0.03}$ & $2.041$ & $2.169$ \\ 
$\sigma_{8,0}$ &$0.8152$ & $0.8191_{-0.011}^{+0.011}$ & $0.7973$ & $0.8412$ \\ 
\hline 
 \end{tabular} \\ 
\caption{The best-fit, mean, $1\sigma$ errors and $95\%$ confidence limits for the $w$CDM model for the data combination CMB+Lensing+BAO+SnIa. In this case the minimum was found for $\chi^2=3810$. \label{tab:wcdm1}}
\end{table}

\begin{table}[!ht]
\begin{tabular}{|l|c|c|c|c|} 
 \hline 
Param & best-fit & mean$\pm\sigma$ & 95\% lower & 95\% upper \\ \hline 
$n_s$ &$0.9692$ & $0.9663_{-0.004}^{+0.0039}$ & $0.9584$ & $0.9742$ \\ 
$w$ &$-1.005$ & $-1.013_{-0.028}^{+0.029}$ & $-1.07$ & $-0.9554$ \\ 
$\Omega_\mathrm{m,0}$ &$0.3061$ & $0.3064_{-0.0082}^{+0.0071}$ & $0.2917$ & $0.3217$ \\ 
$10^{+9} A_s$ &$2.112$ & $2.101_{-0.032}^{+0.03}$ & $2.037$ & $2.164$ \\ 
$\sigma_{8,0}$ &$0.8114$ & $0.8115_{-0.0099}^{+0.01}$ & $0.7911$ & $0.8314$ \\ 
\hline 
 \end{tabular} \\ 
\caption{The best-fit, mean, $1\sigma$ errors and $95\%$ confidence limits for the $w$CDM model for the data combination CMB+Lensing+BAO+SnIa+RSD. In this case the minimum was found for $\chi^2=3826$. \label{tab:wcdm2}}
\end{table}

\begin{table}[!ht]
\begin{tabular}{|l|c|c|c|c|} 
 \hline 
Param & best-fit & mean$\pm\sigma$ & 95\% lower & 95\% upper \\ \hline 
$n_s$ &$0.966$ & $0.9651_{-0.004}^{+0.0038}$ & $0.9571$ & $0.973$ \\ 
$w$ &$-0.9978$ & $-1.027_{-0.027}^{+0.033}$ & $-1.086$ & $-0.9743$ \\ 
$10^{+7}c_0$ &$-0.3492$ & $-0.2056_{-0.400}^{+0.400}$ & $-1.000$ & $1.000$ \\ 
$n$ &$0.4127$ & $0.5019_{-0.12}^{+0.083}$ & $0.200$ & $0.800$ \\ 
$\Omega_\mathrm{m,0}$ &$0.3083$ & $0.3063_{-0.0074}^{+0.0074}$ & $0.2914$ & $0.3211$ \\ 
$10^{+9} A_s$ &$2.098$ & $2.104_{-0.032}^{+0.029}$ & $2.041$ & $2.168$ \\ 
$\sigma_{8,0}$ &$0.8078$ & $0.819_{-0.011}^{+0.01}$ & $0.7983$ & $0.8402$ \\ 
\hline 
 \end{tabular} \\ 
\caption{The best-fit, mean, $1\sigma$ errors and $95\%$ confidence limits for the nonadiabatic model for the data combination CMB+Lensing+BAO+SnIa. In this case the minimum was found for $\chi^2=3809$. \label{tab:nonad1}}
\end{table}

\begin{table}[!ht]
\begin{tabular}{|l|c|c|c|c|} 
 \hline 
Param & best-fit & mean$\pm\sigma$ & 95\% lower & 95\% upper \\ \hline 
$n_s$ &$0.9638$ & $0.9662_{-0.0041}^{+0.004}$ & $0.9582$ & $0.9742$ \\ 
$w$ &$-1.023$ & $-1.016_{-0.027}^{+0.031}$ & $-1.071$ & $-0.9608$ \\ 
$10^{+7}c_0$ &$-0.08274$ & $0.001678_{-0.28}^{+0.36}$ & $-0.7133$ & $0.7427$ \\ 
$n$ &$0.5417$ & $0.4843_{-0.11}^{+0.12}$ & $0.200$ & $0.800$ \\ 
$\Omega_\mathrm{m,0}$ &$0.3041$ & $0.3059_{-0.0074}^{+0.0076}$ & $0.291$ & $0.3206$ \\ 
$10^{+9} A_s$ &$2.097$ & $2.099_{-0.031}^{+0.029}$ & $2.037$ & $2.162$ \\ 
$\sigma_{8,0}$ &$0.8136$ & $0.8122_{-0.01}^{+0.0097}$ & $0.7925$ & $0.832$ \\ 
\hline 
 \end{tabular} \\ 
\caption{The best-fit, mean, $1\sigma$ errors and $95\%$ confidence limits for the nonadiabatic model for the data combination CMB+Lensing+BAO+SnIa+RSD. In this case the minimum was found for $\chi^2=3827$. \label{tab:nonad2}}
\end{table}

\begin{figure*}[!t]
\centering
\includegraphics[width=0.9\textwidth]{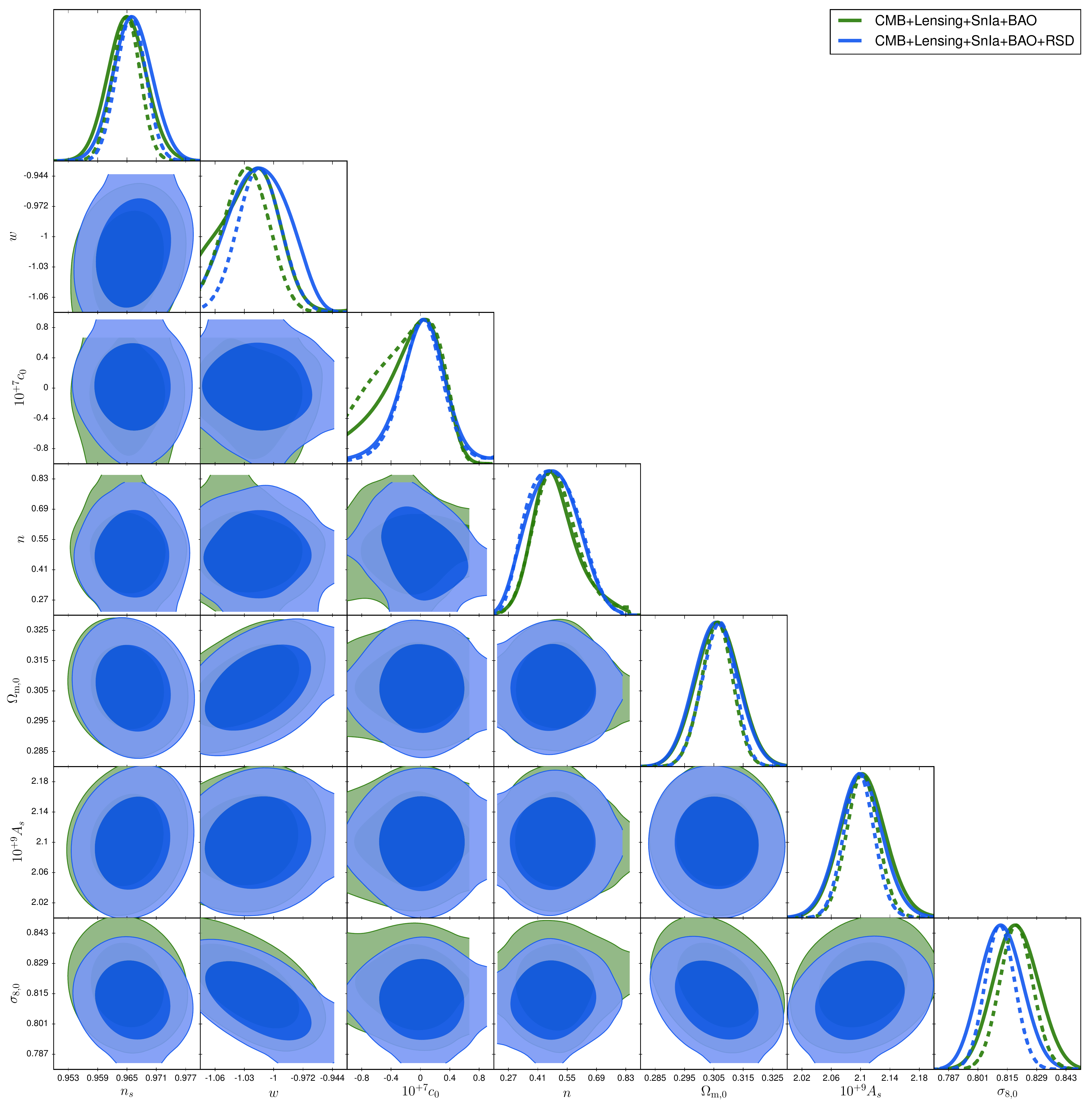}
\caption{The confidence contours for $w$CDM with nonadiabatic DE perturbations using the data combinations of  Planck+Lensing+BAO+SnIa (green contours) and  Planck+Lensing+BAO+SnIa+RSD (blue contours). Some of the contours appear to be truncated due to a peculiarity of the MontePython plotting routines and not due to our choice of the prior. \label{fig:mcmcnonad}}
\end{figure*}

Next, in Fig.~\ref{fig:mcmcnonad} we present the constraints for the nonadiabatic model. In particular we show the confidence contours using CMB+Lensing+BAO+SnIa (green contours) and the CMB+Lensing+BAO+SnIa+RSD (blue contours), while in Tables~\ref{tab:nonad1}-\ref{tab:nonad2} we present the $68\%$ mean values and $95\%$ confidence regions, for some of the parameters of the model. As can be seen, the amplitude of the nonadiabatic perturbation $c_0$ is consistent with zero, while $n$ is very close to $n\sim1/2$ as expected from the toy model based on the $f(R)$ designer model.


\section{Conclusions\label{sec:conclusions}}
In this work we have explored the effects of a nonadiabatic DE pressure perturbation on the CMB and LSS. First, we derived the extra contribution of this nonadiabatic component on the DE perturbation equations, given by the last terms in Eqs.~\eqref{eq:eqd2} and \eqref{eq:eqth2}. Since currently it is unknown if DE has a nonadiabatic component and, even if it does, the behavior of $\frac{\hat{\delta P}_{\textrm{nad}}}{\bar{\rho}}$ is unknown, we took advantage of the effective fluid approach of Refs.~\cite{Arjona:2018jhh}-\cite{Arjona:2019rfn} in order to construct a realistic ansatz. 

In particular, using the designer $f(R)$ model, we derived the expected behavior of this nonadiabatic component both at early and late times, finding that in either era it can be modeled as a power law. Inspired by this, we then assumed the ansatz given by Eq.~\eqref{eq:Pansatz}, where from the $f(R)$ model we expect $n\sim0.5$. We then solved the fluid equations and implemented it into the Boltzmann code CLASS. Moreover, using an approach similar to that of Ref.~\cite{Nesseris:2015fqa}, we were able to find analytical approximations to the growth rate of matter perturbations $f\sigma_8(z)$ of better than $0.5\%$ when compared with our numerical implementation in CLASS.

Since we expect the DE perturbations to have an effect, if at all, at late times when they are growing, we anticipate the nonadiabatic component will affect the CMB only at late times and on large scales. Equivalently, this implies that it affects the low multipoles via the ISW effect and using our implementation in CLASS, we confirmed this. Furthermore, availing ourselves of the modifications in CLASS, we also performed MCMC analyses using the latest cosmological data. Here, we used CMB, BAO and SnIa data, as well as a new RSD likelihood for MontePython, which we presented in this work for the first time. By doing this analysis we found that the parameter $c_0$ is consistent with zero at $1\sigma$, while $n\sim0.5$ is in agreement with the expectation from the designer $f(R)$ model.

In conclusion, we have shown that a nonadiabatic DE pressure perturbation could have measurable effects on the CMB and other key cosmological observables such as the growth rate of matter density perturbations and the matter power spectrum. Using the latest cosmological data, including RSDs, and assuming a power-law for the nonadiabatic DE component given by Eq.~\eqref{eq:Pansatz}, we constrained its amplitude and found it is consistent with zero and GR at $1\sigma$. 

\section*{Acknowledgements}
It is our pleasure to thank G.~Ballesteros,  T.~Brinckmann and W.~Wright for useful discussions. The authors acknowledge use of the CLASS and MontePython codes. Part of the calculations took place at the Hydra HPC Cluster of the  Instituto de F\'isica Te\'orica UAM/CSIC. The authors acknowledge support from the Research Project PGC2018-094773-B-C32 [MINECO-FEDER], and the Centro de Excelencia Severo Ochoa Program SEV-2016-0597. Finally, S.N. also acknowledges support from the Ram\'{o}n y Cajal program through Grant No. RYC-2014-15843. \\

\section*{Numerical codes}
The publicly available RSD Montepython likelihood for the growth rate $f\sigma_8$ data set, introduced in this paper for the first time, is based on the compilation shown in Table~\ref{tab:fs8tab} and can be found at \href{https://github.com/snesseris/RSD-growth}{https://github.com/snesseris/RSD-growth}. 

\begin{appendix} 
\section{The RSD likelihood\label{sec:appendix1}}
Here we describe the RSD likelihood we used for the MCMC analysis done in the previous sections. In particular, we implement in python a likelihood for the ``Gold 2018" growth rate $f\sigma_8$ compilation with $N=22$ data points given in  Ref.~\cite{Sagredo:2018ahx} and shown in Table~\ref{tab:fs8tab} with the corresponding references of each point. 

\begin{table}[!t]
\caption[]{Compilation of the $f\sigma_8(z)$ measurements used in this analysis along with the reference matter density parameter $\Omega_{m_0}$ (needed for the growth correction) and related references.	 \label{tab:fs8tab}}
\begin{center}
\begin{tabular}{ccccccccc}
\hline
\hline
$z$     & $f\sigma_8(z)$ & $\sigma_{f\sigma_8}(z)$  & $\Omega_{m,0}^\text{ref}$ & Ref. \\ \hline
0.02    & 0.428 & 0.0465  & 0.3 & \cite{Huterer:2016uyq}   \\
0.02    & 0.398 & 0.065   & 0.3 & \cite{Turnbull:2011ty},\cite{Hudson:2012gt} \\
0.02    & 0.314 & 0.048   & 0.266 & \cite{Davis:2010sw},\cite{Hudson:2012gt}  \\
0.10    & 0.370 & 0.130   & 0.3 & \cite{Feix:2015dla}  \\
0.15    & 0.490 & 0.145   & 0.31 & \cite{Howlett:2014opa}  \\
0.17    & 0.510 & 0.060   & 0.3 & \cite{Song:2008qt}  \\
0.18    & 0.360 & 0.090   & 0.27 & \cite{Blake:2013nif} \\
0.38    & 0.440 & 0.060   & 0.27 & \cite{Blake:2013nif} \\
0.25    & 0.3512 & 0.0583 & 0.25 & \cite{Samushia:2011cs} \\
0.37    & 0.4602 & 0.0378 & 0.25 & \cite{Samushia:2011cs} \\
0.32    & 0.384 & 0.095  & 0.274 & \cite{Sanchez:2013tga}   \\
0.59    & 0.488  & 0.060 & 0.307115 & \cite{Chuang:2013wga} \\
0.44    & 0.413  & 0.080 & 0.27 & \cite{Blake:2012pj} \\
0.60    & 0.390  & 0.063 & 0.27 & \cite{Blake:2012pj} \\
0.73    & 0.437  & 0.072 & 0.27 & \cite{Blake:2012pj} \\
0.60    & 0.550  & 0.120 & 0.3 & \cite{Pezzotta:2016gbo} \\
0.86    & 0.400  & 0.110 & 0.3 & \cite{Pezzotta:2016gbo} \\
1.40    & 0.482  & 0.116 & 0.27 & \cite{Okumura:2015lvp} \\
0.978   & 0.379  & 0.176 & 0.31 & \cite{Zhao:2018jxv} \\
1.23    & 0.385  & 0.099 & 0.31 & \cite{Zhao:2018jxv} \\
1.526   & 0.342  & 0.070 & 0.31 & \cite{Zhao:2018jxv} \\
1.944   & 0.364  & 0.106 & 0.31 & \cite{Zhao:2018jxv} \\
\hline
\hline
\end{tabular}
\end{center}
\end{table}

The growth data used here are obtained from RSD  measurements, which probe the LSS. In practice they measure the parameter $f\sigma_8(a)\equiv f(a)\cdot \sigma(a)$, where $f(a)=\frac{d ln\delta}{d lna}$ is the growth rate and $\sigma(a)=\sigma_{8,0}\frac{\delta(a)}{\delta(1)}$ denotes the redshift-dependent rms fluctuations of the linear density field within spheres of radius $R=8h^{-1}$Mpc, where by  $\sigma_{8,0}$ we denote its present value. This particular dataset was shown in Ref.~\cite{Sagredo:2018ahx} to be internally robustness and unbiased by using the ``robustness'' criterion of Ref.~\cite{Amendola:2012wc}, by which  combinations of subsets in the dataset underwent a Bayesian analysis and the dataset's overall consistency was established. 

This compilation was also used in Ref.~\cite{Sagredo:2018rvc}, to place constraints on the dark-matter pressure, sound speed and viscosity. Some other compilations also exist in the literature (see Refs.~\cite{Kazantzidis:2018rnb, Kazantzidis:2018jtb, Skara:2019usd}) but these contain duplicate points coming from the same surveys but in different years, as the goal of their analysis was to study the evolution of the $f\sigma_8$ tension over time. Here we will only focus on the compilation given in Table~\ref{tab:fs8tab}, as these points are unique and their statistical robustness has already been confirmed \cite{Sagredo:2018ahx}.

By using the ratio of the monopole to the quadrupole of the redshift-space power spectrum, $f\sigma_8(a)$ can be measured directly and it can be shown that assuming linear theory $f\sigma_8(a)$ is independent of the bias parameter $b(k,z)$, as the latter completely cancels out  \cite{Percival:2008sh,Song:2008qt,Nesseris:2006er}.  Moreover, and more importantly for this analysis, it has been shown that $f\sigma_8(a)$ can also discriminate between DE models \cite{Song:2008qt}. 

In Table~\ref{tab:fs8tab} the RSD data points are given in different redshifts as $ f\sigma_8^\textrm{obs,i}=\Big(f\sigma_8^\textrm{obs}(z_1),\dots, f\sigma_8^\textrm{obs}(z_n)\Big)$, while the theoretical prediction is given by  $f\sigma_8^{\textrm{th}}(\bm \theta_p)=\Big(f\sigma_8^\textrm{th}(z_1),\dots, f\sigma_8^{th}(z_n)\Big)$, which depends on the cosmological model and the parameters $\theta_p$. Note however, that some of the points are correlated with each other, and they also assume a fiducial cosmology that has to be corrected for due to the Alcock-Paczynski effect; see Refs.~\cite{Sagredo:2018ahx,Nesseris:2017vor,Kazantzidis:2018rnb}, and for earlier analyses see Refs.~\cite{Basilakos:2018arq,Basilakos:2017rgc,Basilakos:2016nyg}. We give the values of the $\Omega_{m0}$ parameter for the fiducial flat \lcdm model used in the fourth column of Table.~\ref{tab:fs8tab}. 

The correlated data points are the three WiggleZ points from Ref.~\cite{Blake:2012pj} and the four points from  SDSS \cite{Zhao:2018jxv}. The covariance matrix of the WiggleZ data is given by
\begin{equation}\label{WiggleZCov}
      \mathbf{C}_{\text{WiggleZ}}= 10^{-3}
    \left(
         \begin{array}{ccc}
           6.400 & 2.570 & 0.000 \\
           2.570 & 3.969 & 2.540 \\
           0.000 & 2.540 & 5.184 \\
         \end{array}
       \right),
\end{equation}
while the covariance matrix of the SDSS points is given by
\begin{equation}\label{SDSS4Cov}
     \mathbf{C}_{\text{SDSS-IV}}= 10^{-2}
    \left(
         \begin{array}{cccc}
   3.098 & 0.892 &  0.329 & -0.021\\
       0.892 & 0.980 & 0.436 & 0.076\\
       0.329 & 0.436 &  0.490   & 0.350 \\
       -0.021 & 0.076 & 0.350 & 1.124
       \end{array}
       \right).
\end{equation}

The redshift correction for the Alcock-Paczynski effect as described in Ref.~\cite{Nesseris:2017vor},  is given in terms of  a correction factor of
\begin{equation}
\text{fac}(z^i)= \frac{H(z^i)\,d_A(z^i)}{H^{\text{ref},i}(z^i) \, d_A^{\text{ref},i}(z^i)}\;,
\end{equation}
where the label ``$\text{ref},i$'' stands for the fiducial cosmology used on each data point at the redshift $z^i$. As a result, the now corrected growth rate is \cite{Macaulay:2013swa}
\begin{equation}
f\sigma_8^{\textrm{th,i}}\rightarrow\frac{f\sigma_8^{\textrm{th,i}}}{\text{fac}(z^i)}\;.
\end{equation}
We can then define the data vector $\bm V$ as:
\begin{equation}
     \mathbf{V} =  \mathbf{f\sigma_8^\textrm{obs,i}} -  \frac{f\sigma_8^{\textrm{th,i}}}{\text{fac}(z^i)},
\end{equation}
and the chi-squared of our likelihood via
\begin{equation}
	\chi^2=  \mathbf{x}^T  \mathbf{C}^{-1}  \mathbf{x}\;.
\end{equation}
Finally, in CLASS we can obtain the scale-dependent growth $\delta(k,z)$ at each redshift via the matter power spectrum as $\delta(k,z)=\sqrt{\frac{P(k,z)}{P(k,0)}}$, where the matter power spectrum $P(k,z)$ is obtained from the code itself via the function \texttt{cosmo.pk(k,z)}. Then, $f\sigma_8(k,z)$ can be obtained with simple cubic interpolations and direct differentiation from Eq.~\eqref{eq:fs8def}.

\end{appendix}

\bibliography{na_cs}

\end{document}